\documentclass[review]{elsarticle}
\usepackage{amsmath, amssymb, eqparbox}
\usepackage{graphicx,psfrag,epsf}
\usepackage{enumerate}
\usepackage{url} 
\usepackage[margin=20mm]{geometry}
\usepackage{algorithm}
\usepackage{algpseudocode}
\usepackage{amsthm}
\usepackage{mathrsfs}
\usepackage{multirow}
\usepackage{pdfpages}
\usepackage{hyperref}
\usepackage{float}

\journal{Expert Systems with Applications}

\biboptions{authoryear}

\usepackage[hang,small,bf]{caption}
\usepackage[subrefformat=parens]{subcaption}
\theoremstyle{plain}
\newtheorem{thm}{Theorem}
\newtheorem*{thm*}{Theorem}

\theoremstyle{definition}
\newtheorem{dfn}{Definition}

\DeclareMathOperator*{\argmin}{arg\,min}

\floatstyle{plaintop}
\restylefloat{table}

\makeatletter
\def\ps@pprintTitle{%
 \let\@oddhead\@empty
 \let\@evenhead\@empty
 \let\@oddfoot\@empty
 \let\@evenfoot\@empty
}
\makeatother

\begin{document}
\begin{frontmatter}

\begin{titlepage}
\begin{center}
\vspace*{1cm}

\textbf{ \large Estimation of conditional average treatment effects on distributed confidential data}

\vspace{1.5cm}

Yuji Kawamata$^{a}$ (yjkawamata@gmail.com), Ryoki Motai$^b$ (motai.ryoki.tg@alumni.tsukuba.ac.jp), Yukihiko Okada$^{a,c}$ (okayu@sk.tsukuba.ac.jp), Akira Imakura$^{a,c}$ (imakura@cs.tsukuba.ac.jp), Tetsuya Sakurai$^{a,c}$ (sakurai@cs.tsukuba.ac.jp) \\

\hspace{10pt}

\begin{flushleft}
\small  
$^a$ Center for Artificial Intelligence Research, Tsukuba Institute for Advanced Research, University of Tsukuba, 1-1-1 Tennodai, Tsukuba, Ibaraki 305-8577, Japan \\
$^b$ Graduate School of Science and Technology, University of Tsukuba, 1-1-1 Tennodai, Tsukuba, Ibaraki 305-8573, Japan \\
$^c$ Faculty of Engineering, Information and Systems, University of Tsukuba, 1-1-1 Tennodai, Tsukuba, Ibaraki 305-8573, Japan

\vspace{1cm}
\textbf{Corresponding Author:} \\
Yuji Kawamata \\
Center for Artificial Intelligence Research, Tsukuba Institute for Advanced Research, University of Tsukuba, 1-1-1 Tennodai, Tsukuba, Ibaraki 305-8577, Japan \\
Email: yjkawamata@gmail.com

\end{flushleft}        
\end{center}
\end{titlepage}

\title{Estimation of conditional average treatment effects on distributed confidential data}

\author[inst_cair]{Yuji Kawamata\corref{cor1}}
\ead{yjkawamata@gmail.com}

\author[inst_gs]{Ryoki Motai}
\ead{motai.ryoki.tg@alumni.tsukuba.ac.jp}

\author[inst_cair,inst_feis]{Yukihiko Okada}
\ead{okayu@sk.tsukuba.ac.jp}

\author[inst_cair,inst_feis]{Akira Imakura}
\ead{imakura@cs.tsukuba.ac.jp}

\author[inst_cair,inst_feis]{Tetsuya Sakurai}
\ead{sakurai@cs.tsukuba.ac.jp}

\cortext[cor1]{Corresponding author.}
\address[inst_cair]{Center for Artificial Intelligence Research, Tsukuba Institute for Advanced Research, University of Tsukuba, 1-1-1 Tennodai, Tsukuba, Ibaraki 305-8577, Japan}
\address[inst_gs]{Graduate School of Science and Technology, University of Tsukuba, 1-1-1 Tennodai, Tsukuba, Ibaraki 305-8573, Japan}
\address[inst_feis]{Faculty of Engineering, Information and Systems, University of Tsukuba, 1-1-1 Tennodai, Tsukuba, Ibaraki 305-8573, Japan}

\begin{abstract}
The estimation of conditional average treatment effects (CATEs) is an important topic in many scientific fields.
CATEs can be estimated with high accuracy if data distributed across multiple parties are centralized.
However, it is difficult to aggregate such data owing to confidentiality or privacy concerns.
To address this issue, we propose data collaboration double machine learning, a method for estimating CATE models using privacy-preserving fusion data constructed from distributed sources, and evaluate its performance through simulations.
We make three main contributions.
First, our method enables estimation and testing of semi-parametric CATE models without iterative communication on distributed data, providing robustness to model mis-specification compared to parametric approaches.
Second, it enables collaborative estimation across different time points and parties by accumulating a knowledge base.
Third, our method performs as well as or better than existing methods in simulations using synthetic, semi-synthetic, and real-world datasets.
\end{abstract}




\end{frontmatter}

\section{Introduction}
\label{sec:intro}

The Neyman--Rubin model \citep{imbens2015causal} or the potential outcomes framework is one of the main causal inference methods to estimate the average effects of treatment (intervention) and has been developed and applied in many studies since it was established by Rubin.
In recent years, there have been innovations that allow the estimation of individual or conditional average treatment effects (CATEs) by adapting not only statistical but also machine learning methods (e.g. \cite{kunzel2019metalearners, athey2019generalized}).
Most methods for estimating CATEs assume that the data can be centralized in one place.
However, if the distributed data contain confidential or private information, it is difficult to centralize them in one place.

Conversely, in the field of machine learning, in recent years, a privacy-preserving analysis method called federated learning has been developed \citep{konevcny2016federated, mcmahan2017communication, rodriguez2020federated, criado2022non, rodriguez2023survey, rafi2024fairness}.
Data collaboration (DC) analysis \citep{imakura2020data, Bogdanova2020federated, imakura2021interpretable, imakura2023dccox, imakura2023non}, one of federated learning systems, is a method that enables collaborative analysis from privacy-preserving fusion data.
The privacy-preserving fusion data are constructed by sharing dimensionality-reduced intermediate representations instead of raw data.
DC analysis was originally proposed to address regression and classification problems in distributed data.
Recently, Kawamata et al. \citep{kawamata2024collaborative} proposed data collaboration quasi-experiment (DC-QE) that extends DC analysis to allow estimation of average treatment effect (ATE).
It is important to note that although those studies often focus their discussions on privacy data (personal information), the technologies can also be used to preserve the management information held by companies. Therefore, in this paper, we collectively refer to such data that are difficult to disclose to the outside as confidential information. Below we discuss techniques for preserving confidential information.

In this paper, we propose data collaboration double machine learning (DC-DML) as a method to estimate CATEs while preserving the confidentiality of distributed data by incorporating double machine learning (DML) \citep{chernozhukov2018double}, a machine learning-based treatment effect estimation method, into DC-QE.
Moreover, through numerical simulations, we evaluate the performance of our method compared to existing methods.

This paper builds on partially linear models, one of the semi-parametric models.
Since Engle et al. \citep{engle1986semiparametric}, partially linear models have been used in empirical studies where assuming a linear model between the covariates and the outcome is not appropriate, i.e., to avoid model mis-specification owing to linear assumptions.
A significant advance in partially linear models in recent years is DML \citep{chernozhukov2018double}, which brought a least square estimator that can adapt common machine learning models.
Various causal inference methods have been developed by developing DML \citep{fan2022estimation, bia2023double}.
In this paper, we develop DML into a method that enables CATE estimation through distributed data with privacy-preserving.

The effects of policy interventions by governments or medical treatments by hospitals are likely to differ across people.
It is possible to improve the overall performance of the treatment if one knows which subjects should be treated.
For example, by concentrating scholarship support on those whose future income increases will be higher owing to the treatment, the limited budget can be used more efficiently.
In addition, by concentrating medication on those who are more likely to improve or have fewer side effects from the treatment, medication can be more effective and safer to use.
One approach to estimating CATEs for these situations with a high accuracy is to centralize data from many parties, but this is difficult owing to confidential or privacy concerns.
Our method enables estimation of CATEs with high accuracy while maintaining the confidentiality of distributed data across many parties.

\begin{figure}[tb]
  \centering
  \includegraphics[width=0.8\linewidth]{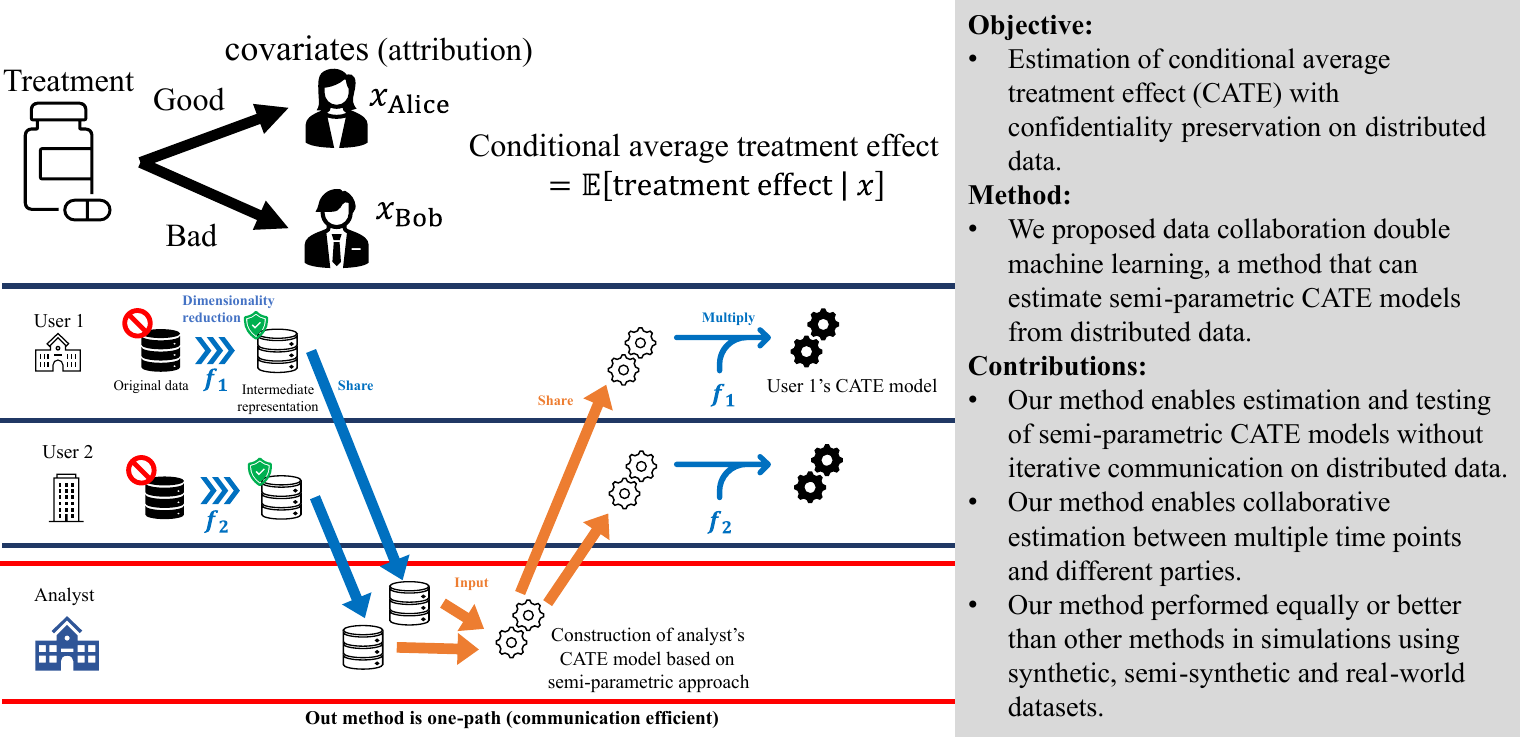}
  \caption{The summary of our study.}
  \label{fig:graphical_abstract}
\end{figure}

Our contributions are summarized in the following three points.
First, our method enables estimation and testing of semi-parametric CATE models without iterative communication on distributed data.
Semi-parametric (or non-parametric) CATE models enable estimation and testing that is more robust to model mis-specification than parametric models.
However, to our knowledge, no communication-efficient method has been proposed for estimating and testing semi-parametric (or non-parametric) CATE models on distributed data.
Second, our method enables collaborative estimation between different parties as well as multiple time points because the dimensionality-reduced intermediate representations can be accumulated as a knowledge base.
Third, our method performs as well or better than other methods in evaluation simulations using synthetic, semi-synthetic and real-world datasets.
Fig. \ref{fig:graphical_abstract} shows the summary of our study.

The remainder of this paper is organized as follows.
In Section \ref{sec:relatedwork}, we present related works on causal inference and machine learning.
We describe the basics of causal inference and distributed data in Section \ref{sec:linear_DML}.
We describe our method in Section \ref{sec:method}.
In Section \ref{sec:simulation}, we report the numerical simulations.
In Section \ref{sec:discussion}, we discuss the results.
Finally, we conclude in Section \ref{sec:conclusion}.
Acronyms and their definitions used in this paper are summarized in \ref{app:acronym}.

\section{Related works}
\label{sec:relatedwork}
The field of treatment effect estimation has advanced significantly in recent years through the incorporation of machine learning methods as well as statistics.
However, studies that take into account confidentiality preservation, which is our aim, are limited.
Here, we briefly review existing methods.

DML is an innovative method proposed by Chernozhukov et al. \cite{chernozhukov2018double} that allows estimation of ATE using any machine learning method.
DML is a semi-parametric method that uses machine learning to estimate nuisance parameters for estimating target parameters.
More specifically, DML first estimates nuisance parameters using machine learning, and then estimates the target parameters using the estimated nuisance parameters.
DML resolves influences of regularization bias and over-fitting on estimates of target parameters through Neyman--orthogonal scores and cross-fitting, respectively.
As shown in Section \ref{sec:linear_DML}, DML can be easily extended to allow estimation of CATE.
DML is a useful method for estimating treatment effects, but using centralized data.
To achieve our goal, we need to develop DML to address distributed data.
In Section \ref{sec:linear_DML}, we describe DML in detail.

As representative methods that utilize machine learning for CATE estimation, generalized random forest (GRF) \citep{athey2019generalized} and X-Learner \citep{kunzel2019metalearners} have been proposed, but these methods cannot be applied to distributed data.
\footnote{
Many other methods for estimating CATEs have been proposed so far, such as non-parametric Bayesian \citep{hill2011bayesian, taddy2016nonparametric} and model averaging \citep{rolling2019combining, shi2024estimating}.
However, these methods cannot also be applied to distributed data.
}
GRF is a non-parametric method for estimating treatment effects by solving local moment equations using random forests.
GRF first constructs a random forest, then calculates weights for each sample from the random forest, and finally derives the treatment effect by solving a moment equations using the calculated weights.
X-Learner is a non-parametric method that can estimate treatment effects from any machine learning model.
X-Learner uses propensity scores \citep{rosenbaum1983central} to resolve performance degradation that occurs when controlled or treated group is much larger than the other group.

Conversely, Secure regression (SR) \citep{karr2005secure}, FedCI \citep{vo2022bayesian} and CausalRFF \citep{vo2022adaptive} are methods that can estimate CATEs from distributed data.
SR is a parametric method for estimating linear regression models.
In SR, statistics necessary to compute least squares estimates are calculated in each local party and these statistics are shared.
SR is communication-efficient because it enables the estimation of a linear model by sharing statistics only once.
However, in SR, bias owing to mis-specification can occur if the assumptions of the linear model are incorrect.
FedCI is a non-parametric method that constructs a potential outcome model based on Gaussian processes through the federated learning approach.
In FedCI, the communication between the clients and the server is iterated: each client sends the model gradient calculated from its local data to the server, then the server averages the aggregated gradients, updates its own model parameters, and returns those parameters to the clients.
In the case where local data are stored in a location isolated from the Internet or communication synchronization is not possible, FedCI is difficult to execute.
CausalRFF \citep{vo2022adaptive} is a kernel-based method designed to estimate CATEs under latent confounding in distributed data settings.
It leverages random Fourier features to model treatment effects.
Although our study does not consider latent confounders, both CausalRFF and our method exploit latent space representations constructed from distributed data.

While the purpose of our method is to estimate CATEs, ifedtree \citep{tan2022tree}, Federated MLE, IPW-MLE and AIPW \citep{xiong2023federated}, and DC-QE \citep{kawamata2024collaborative} are methods that have been proposed for different estimation targets.
Those methods can address distributed data.
ifedtree is a tree-based model averaging approach to improve the CATE estimation at a specific party by sharing models derived from other parties instead of raw data.
Our method aims to estimate CATEs on aggregated whole data, rather than on specific party data.
Federated MLE, IPW-MLE and AIPW are parametric methods that aggregate individually estimated gradients for estimating propensity scores and treatment effects.
Federated MLE, IPW-MLE and AIPW estimate the ATE not conditioned on covariates.

DC-QE is a semi-parametric method for estimating ATE based on a DC analysis framework.
DC-QE first construct collaborative representations based on dimensionality-reduced intermediate representations collected from all parties instead of raw data, second estimates propensity scores, finally estimates ATE using the estimated propensity scores (see \ref{app:alg_deqe} for the pseudo-code of DC-QE).
Our method extends DC-QE to address DML for CATE estimation.

Table \ref{tab:methods} summarizes six of those described in this section that are particularly relevant to our method.
In Section \ref{sec:method}, we propose a method for estimating CATEs in a semi-parametric model while preserving confidentiality of distributed data.
Although DML, GRF and X-Learner are methods for centralized data, our method addresses distributed data.
Moreover, our method is more robust than SR for model mis-specification and more communication-efficient than FedCI and CausalRFF.

\begin{table}[tb]
\centering
    \scalebox{0.8}{
        \begin{tabular}{lccc}
        \hline \hline
        Methods & \multicolumn{1}{l}{\begin{tabular}[c]{@{}l@{}}Can it deal   with\\ distributed data?\end{tabular}} & \multicolumn{1}{l}{\begin{tabular}[c]{@{}l@{}}Assumptions of\\ CATE model\end{tabular}} & \multicolumn{1}{l}{\begin{tabular}[c]{@{}l@{}}Does it need\\ iterative communication?\end{tabular}} \\ \hline
        DML \citep{chernozhukov2018double} & No & Semi-parametric & - \\ \hline
        GRF \citep{athey2019generalized} & No & Non-parametric & - \\ \hline
        X-Learner \citep{kunzel2019metalearners} & No & Non-parametric & - \\ \hline
        SR \citep{karr2005secure} & Yes & Parametric & No \\ \hline
        CausalRFF \citep{vo2022adaptive} & Yes & Semi-parametric & Yes \\ \hline
        FedCI \citep{vo2022bayesian} & Yes & Non-parametric & Yes \\ \hline
        \textbf{DC-DML (our method)} & \textbf{Yes} & \textbf{Semi-parametric} & \textbf{No} \\ \hline \hline
        \end{tabular}
    }
\caption{Existing and our methods.}
\label{tab:methods}
\end{table}

\section{DML with the linear CATE model}
\label{sec:linear_DML}
In this section, we describe the DML method with the linear CATE model.
We adopt the Neyman--Rubin potential outcomes framework, in which we define the following variables for each subject $i$: $y_i(1)$ and $y_i(0)$ denote the potential outcomes under treatment and control, respectively; $z_i \in \{0,1\}$ denotes the binary treatment indicator; and $\boldsymbol{x}_i = (x_i^1,\cdots,x_i^m)^\top \in \mathbb{R}^{m}$ is the covariate vector.
Here, $m$ is the number of covariates.
The observed outcome is denoted by \( y_i = y_i(z_i) \).

Under this framework, we make the following standard assumptions \citep{imbens2015causal, hernan2016does} required for causal identification of the CATE:
\begin{itemize}
    \item Stable unit treatment value assumption: There is no interference between subjects, and each treatment has a single version. That is, $y_i(z)$ depends only on the treatment assigned to subject $i$.
    \item Consistency: The observed outcome corresponds to the potential outcome under the received treatment, i.e.,  $y_i = y_i(z_i)$.
    \item Unconfoundedness: Given covariates $\boldsymbol{x}_i$, the treatment is independent of the potential outcomes: 
        \begin{equation*}
            \left( y_i(1), y_i(0) \right) \perp\!\!\!\perp z_i \mid \boldsymbol{x}_i,
        \end{equation*}
    where $\perp\!\!\!\perp$ denotes statistical independence.
    \item Overlap: For all values of $\boldsymbol{x}_i$, the probability of receiving either treatment is strictly positive:
        \begin{equation*}
            0 < \text{Pr}(z_i = 1 \mid \boldsymbol{x}_i) < 1.
        \end{equation*}
\end{itemize}
Under these assumptions, the CATE is identified as
\begin{equation*}
    \tau(\boldsymbol{x}_i) = \mathbb{E}[ y_i (1) - y_i (0) | \boldsymbol{x}_i].
\end{equation*}

In DML, we assume the following structural equations for the data-generating process:
\begin{align}
    y_i &= \theta(\boldsymbol{x}_i) \cdot z_i + u(\boldsymbol{x}_i) + \varepsilon_i, \label{eq:y} \\
    z_i &= h(\boldsymbol{x}_i) + \eta_i. \label{eq:z}
\end{align}
$\theta$ is a function that represents the effect of the treatment conditional on the covariate $\boldsymbol{x}_i$.
The covariates $\boldsymbol{x}_i$ affect the treatment $z_i$ via the function $h$ and the outcome via the function $u$, where $h$ and $u$ are defined as functions belonging to convex subsets $\mathbb{H}$ and $\mathbb{U}$ with some normed vector space.
$\varepsilon_i$ and $\eta_i$ are disturbances such that $\mathbb{E}[\eta_i |\boldsymbol{x}_i]=0$, $\mathbb{E}[\varepsilon_i|\boldsymbol{x}_i]=0$ and $\mathbb{E}[\eta_i \varepsilon_i|\boldsymbol{x}_i]=0$.
In this data-generating process, the function of our interest is $\theta$ because obviously $\tau(\boldsymbol{x}_i)=\theta(\boldsymbol{x}_i)$, while $h$ and $u$ are not of primary interest.
Note that Chernozhukov et al. \citep{chernozhukov2018double} assumed that $\theta$ is constant, i.e., independent of $\boldsymbol{x}_i$.
However, as we discuss below, $\theta$ can be estimated in a manner similar to Chernozhukov et al. \citep{chernozhukov2018double} even when $\theta$ linearly depends on $\boldsymbol{x}_i$.

In this paper, we derive $\theta$ by a partialling-out approach as follows:
By subtracting the conditional expectation of (\ref{eq:y}) from (\ref{eq:y}), we can rewrite the structural equation as
\begin{equation*}
    y_i - \mathbb{E}[y_i|\boldsymbol{x}_i] = \theta(\boldsymbol{x}_i)\cdot (z_i - \mathbb{E}[z_i|\boldsymbol{x}_i]) + \varepsilon_i.
\end{equation*}
Let $q(\boldsymbol{x}_i) = \mathbb{E}[y_i|\boldsymbol{x}_i]$, where $q$ belongs to a convex subset $\mathbb{Q}$ with some normed vector space.
Then, the residuals for $q$ and $h$ are
\begin{align*}
    \zeta_i &= y_i - q(\boldsymbol{x}_i), \\
    \eta_i &= z_i - h(\boldsymbol{x}_i).
\end{align*}
So we can obtain
\begin{equation}
    \zeta_i = \theta(\boldsymbol{x}_i)\cdot \eta_i + \varepsilon_i.
    \label{eq:res_on_res}
\end{equation}
Note that $\mathbb{E}[\zeta_i | \boldsymbol{x}_i] = 0$ because $\mathbb{E}[\eta_i |\boldsymbol{x}_i]=0$ and $\mathbb{E}[\varepsilon_i|\boldsymbol{x}_i]=0$.

We assume $\theta$ as a linear function with a constant term such that
\begin{equation*}
    \theta(\boldsymbol{x_i}) = [1, \boldsymbol{x}_i^\top] \boldsymbol{\beta} = \beta^0 + x_i^1 \beta^1 + \cdots + x_i^m \beta^m,
\end{equation*}
where $\boldsymbol{\beta} = [\beta^0, \cdots, \beta^m]^\top \in \mathbb{R}^{m+1}$.
Rewrite (\ref{eq:res_on_res}) as
\begin{equation*}
    \zeta_i = \eta_i [1, \boldsymbol{x}_i^\top] \boldsymbol{\beta} + \varepsilon_i.
    \label{eq:res_on_res_cov}
\end{equation*}

To derive $\boldsymbol{\beta}$, consider the following score vector:
\begin{equation}
    \psi(\boldsymbol{x}_i; \hat{\boldsymbol{\beta}}, \hat{q}, \hat{h}) = [1, \boldsymbol{x}_i^\top]^\top (z_i - \hat{h}(\boldsymbol{x}_i))(y_i - \hat{q}(\boldsymbol{x}_i) - [1, \boldsymbol{x}_i^\top] \hat{\boldsymbol{\beta}} (z_i - \hat{h}(\boldsymbol{x}_i))).
    \label{eq:score_vec}
\end{equation}
Define the moment condition
\begin{equation}
    \mathbb{E} [\psi(\boldsymbol{x}_i; \boldsymbol{\beta}, q, h)] = \boldsymbol{0},
    \label{eq:moment}
\end{equation}
and Neyman--orthogonality \citep{chernozhukov2018double}
\begin{equation}
    \lim_{r \to 0} \frac{\partial}{\partial r} \mathbb{E} [\psi(\boldsymbol{x}_i; \boldsymbol{\beta}, q + r(q'-q), h + r(h'-h))] = \boldsymbol{0},
    \label{eq:ney_orth}
\end{equation}
where $r \in [0,1)$, $q' \in \mathbb{Q}$ and $h' \in \mathbb{H}$.
The Neyman--orthogonality implies first-order insensitivity to perturbations of the nuisance functions $q$ and $h$.
Then, we have Theorem \ref{thm:score}, whose proof is given in \ref{app:proof_score}.
\begin{thm}
    The score vector (\ref{eq:score_vec}) satisfies both the moment condition (\ref{eq:moment}) and the Neyman--orthogonality (\ref{eq:ney_orth}).
    \label{thm:score}
\end{thm}

Given $\hat{q}$ and $\hat{h}$, $\hat{\beta}$ can be obtained by solving (\ref{eq:moment}) empirically from the dataset as:
\begin{equation}
    \sum_{i} \psi(\boldsymbol{x}_i; \hat{\boldsymbol{\beta}}, \hat{q}, \hat{h}) = \sum_{i} \left( \hat{\eta_i} [1, \boldsymbol{x}_i^\top]^\top ( \hat{\zeta_i} - \hat{\eta_i} [1, \boldsymbol{x}_i^\top] \hat{\boldsymbol{\beta}}) \right) = \boldsymbol{0}.
    \label{eq:moment_empirically}
\end{equation}
(\ref{eq:moment_empirically}) indicates that $\hat{\boldsymbol{\beta}}$ is the solution to the regression problem from $\hat{\zeta_i}$ to $\hat{\eta_i} [1, \boldsymbol{x}_i^\top]$.
That is,
\begin{equation*}
    \hat{\boldsymbol{\beta}} = \argmin_{\boldsymbol{\beta}'} \left( \sum_{i} \left( \hat{\zeta_i} - \hat{\eta_i} [1, \boldsymbol{x}_i^\top] \boldsymbol{\beta}' \right)^2 \right).
\end{equation*}
DML uses machine learning models to estimate $q$ and $h$, and cross-fitting to reduce the effect of over-fitting on the estimation of the parameter $\theta$.
In this paper, the data are split into two folds in cross-fitting.
To derive the variance of $\hat{\boldsymbol{\beta}}$, rewrite (\ref{eq:score_vec}) as
\begin{equation*}
    \psi(\boldsymbol{x}_i; \hat{\boldsymbol{\beta}}, \hat{q}, \hat{h}) = \psi^a (\boldsymbol{x}_i; \hat{\boldsymbol{\beta}}, \hat{q}, \hat{h}) \hat{\boldsymbol{\beta}} + [1, \boldsymbol{x}_i^\top]^\top (z_i - \hat{h}(\boldsymbol{x}_i))(y_i - \hat{q}(\boldsymbol{x}_i)),
\end{equation*}
where
\begin{equation*}
    \psi^a (\boldsymbol{x}_i; \hat{h}) = (z_i - \hat{h}(\boldsymbol{x}_i))^2 [1, \boldsymbol{x}_i^\top]^\top [1, \boldsymbol{x}_i^\top].
\end{equation*}
From Theorem 3.2 in Chernozhukov et al. \citep{chernozhukov2018double}, the asymptotic variance of $\sqrt{n}(\hat{\boldsymbol{\beta}} - \boldsymbol{\beta})$ is
\begin{equation}
    \hat{\sigma}^2 = \hat{J}^{-1} \frac{1}{K} \sum_{l=1}^L \frac{1}{|N_l|} \sum_{i \in N_l} \psi(\boldsymbol{x}_i; \hat{\boldsymbol{\beta}}, \hat{q}_l, \hat{h}_l) \psi(\boldsymbol{x}_i; \hat{\boldsymbol{\beta}}, \hat{q}_l, \hat{h}_l)^\top (\hat{J}^{-1})^\top
    \label{eq:sigma_hat_2}
\end{equation}
Here, $\hat{q}_l$ and $\hat{h}_l$ are the functions, and $N_l$ is the set of subjects for fold $l$.
Moreover,
\begin{equation}
    \hat{J} = \frac{1}{L} \sum_{l=1}^L \frac{1}{|N_l|} \sum_{i \in N_l} \psi^a (\boldsymbol{x}_i; \hat{h}_l). \label{eq:J_hat}
\end{equation}
Then, the variance of $\hat{\boldsymbol{\beta}}$ is 
\begin{equation}
    \textup{Var}(\hat{\boldsymbol{\beta}}) = \hat{\sigma}^2/n.
    \label{eq:beta_var}
\end{equation}

In summary, the DML procedure can be expressed as the following two steps.
First, $q$ and $h$ are estimated from the two folds using machine learning models to obtain the estimated residuals: $\hat{\zeta}_i$ and $\hat{\eta}_i$.
Second, $\boldsymbol{\beta}$ is estimated by regressing $\hat{\zeta}_i$ to $\hat{\eta}_i [1, \boldsymbol{x}_i^\top]$.

\section{Method}
\label{sec:method}

In this section, we describe our method.
Before that, we explain the data distribution that our method focuses on.
Let $n$ be the number of subjects (sample size) in a dataset.
Moreover, let $ X = [ \boldsymbol{x}_1 , \cdots , \boldsymbol{x}_n ]^\top \in \mathbb{R}^{n \times m}$,
$ Z = [ z_1, \cdots , z_n ]^\top \in \{ 0,1 \}^n$ and
$ Y = [ y_1, \cdots , y_n ]^\top \in \mathbb{R}^{n} $ be the datasets of covariates, treatments, and outcomes, respectively.
We consider that $n$ subjects are partitioned horizontally into $c$ parties, as follows:
\begin{equation*}
    X = \left[
            \begin{array}{c}
            X_1\\ 
            \vdots \\
            X_c
            \end{array}
        \right], \quad
    Z = \left[
            \begin{array}{c}
            Z_1\\ 
            \vdots \\
            Z_c
            \end{array}
        \right], \quad
    Y = \left[
            \begin{array}{c}
            Y_1\\ 
            \vdots \\
            Y_c
            \end{array}
        \right].
\end{equation*}
The $k$th party has a partial dataset and corresponding treatments and outcomes, which are
$X_k \in \mathbb{R}^{n_k \times m}, Z_k \in \{ 0,1 \}^{n_k}, Y_k \in \mathbb{R}^{n_k}$,
where $n_k$ is the number of subjects for each party in the $k$th row ($n=\sum_{k=1}^c n_k$).

This section is organized as follows: we describe our method in Section \ref{sec:method_method}, 
analyze the correctness of our method in Section \ref{sec:method_correctness},
propose a dimension reduction method for DC-DML in Section \ref{sec:method_dimmethod},
discuss the confidentiality preservation of our method in Section \ref{sec:method_conf_prsv},
and finally discuss the advantages and disadvantages in Section \ref{sec:method_ad_disad}.

\subsection{DC-DML}
\label{sec:method_method}
\begin{figure}[tb]
  \centering
  \includegraphics[width=0.5\linewidth]{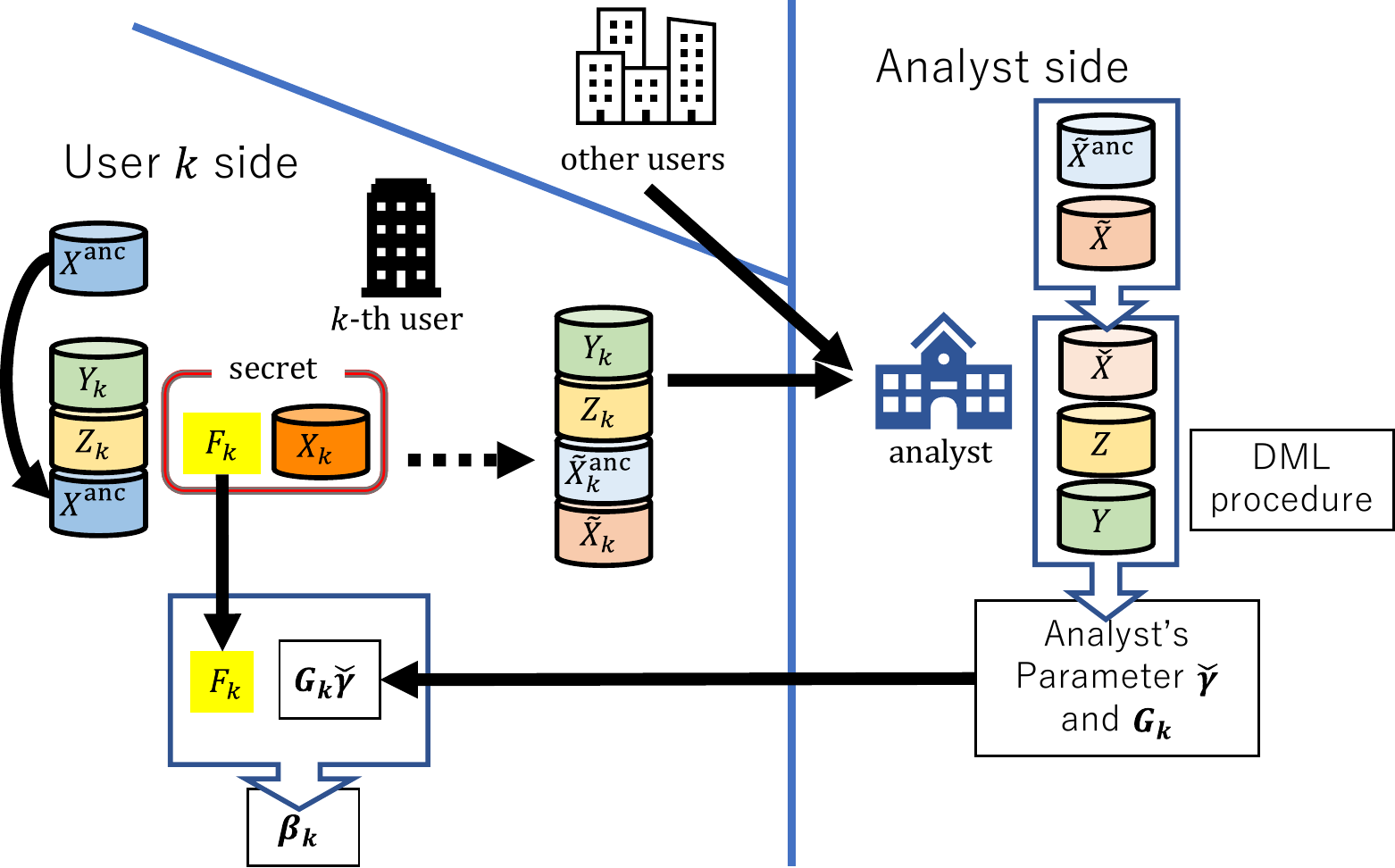}
  \caption{Overall illustration of the proposed method.}
  \label{fig:method}
\end{figure}

Our method DC-DML is based on DC-QE \citep{kawamata2024collaborative} and DML \citep{chernozhukov2018double}, and allows the estimation of CATE while preserving confidentiality of the distributed data.
As in Kawamata et al. \citep{kawamata2024collaborative}, DC-QE operates in two roles: user and analyst roles.
Each user $k$ has confidential covariates $X_k$, treatments $Z_k$ and outcomes $Y_k$.
DC-QE is conducted in three stages: construction of collaborative representations, estimation by the analyst, and estimation by the users.
In the first stage, each user individually constructs intermediate representations and shares them with the analyst instead of the confidential dataset, then the analyst constructs the collaborative representations.
In the second stage, the analyst estimates its own coefficients and variance using DML, and shares them with all users.
In the final stage, each user calculates its own coefficients and variance.
Each stage is described in turn from Section \ref{sec:method_method_construction} to \ref{sec:method_method_estu}.
Fig. \ref{fig:method} shows an overall illustration of DC-DML.
Algorithm \ref{alg:dcdml} is the pseudo-code of DC-DML.

We adopt DML for the analyst-side estimation because it enables not only the estimation of CATEs but also variance estimation for statistical testing, which is critical for causal inference tasks.
X-Learner \citep{kunzel2019metalearners} and GRF \citep{athey2019generalized}, which are, to the best of our knowledge, the most representative non-parametric methods for CATE estimation, do not readily support inference under distributed settings.
In contrast, the semi-parametric structure of DML, which incorporates a linear component, enables each user in the DC framework to test the significance of their CATE models based on formal statistical inference using the returned variance.
This design choice is motivated in part by the related work \citep{imakura2023dccox}, which applied a DC framework to the Cox proportional hazards model, a semi-parametric model.
We acknowledge that other estimation methods may also offer advantages over DML in certain settings; however, extending the DC framework to accommodate such methods is nontrivial and remains to be explored. This presents an important direction for future research.

\begin{algorithm}[tb]
    \caption{Data collaboration double machine learning (DC-DML)}
    \label{alg:dcdml}
    \begin{algorithmic}[1]
        \Statex \textbf{Input: covariates $X$, treatments $Z$, outcomes $Y$}.
        \Statex \textbf{Output: $\boldsymbol{\beta}_k$, $\text{Var}(\boldsymbol{\gamma}_k)$ and $\text{Var}(\alpha_k)$}.
        \vspace{-.2\baselineskip}
        
        \Statex \hrulefill
        \vspace{-.3\baselineskip}
        \Statex \textit{user-$k$-side}
        \vspace{-.5\baselineskip}
        \Statex \hrulefill
        \vspace{-.1\baselineskip}
        \State Generate anchor dataset $X_{k}^\text{anc}$ and share it with all users.
        \State Set $X^\text{anc}$.
        \State Generate $F_{k}$ and $\boldsymbol{\mu}_k$.
        \State Compute $\widetilde{X}_{k} = (X_{k} - \boldsymbol{1} \boldsymbol{\mu}_k^\top ) F_{k}$.
        \State Compute $\widetilde{X}_{k}^\text{anc} = (X^\text{anc} - \boldsymbol{1} \boldsymbol{\mu}_k^\top) F_{k}$.
        \State Share $[\boldsymbol{1} , \widetilde{X}_{k}]$, $[\boldsymbol{1} , \widetilde{X}_{k}^\text{anc}]$, $Z_k$ and $Y_k$ to the analyst.
        \vspace{-.3\baselineskip}
        
        \Statex \hrulefill
        \vspace{-.3\baselineskip}
        \Statex \textit{analyst-side}
        \vspace{-.5\baselineskip}
        \Statex \hrulefill
        \vspace{-.1\baselineskip}
        \State Get $[\boldsymbol{1} , \widetilde{X}_{k}]$, $[\boldsymbol{1} , \widetilde{X}_{k}^\text{anc}]$, $Z_k$ and $Y_k$ for all $k$.
        \State Set $[\boldsymbol{1} , \widetilde{X}_{k}]$ and $[\boldsymbol{1} , \widetilde{X}_{k}^\text{anc}]$.
        \State Compute $G_k$ from $\widetilde{X}_{k}^\text{anc}$ for all $k$ such that $[\boldsymbol{1} , \widetilde{X}_{k}^\text{anc}] G_k \approx [\boldsymbol{1} , \widetilde{X}_{k'}^\text{anc}] G_{k'} ~ (k \neq k')$.
        \State Compute $\check{X}_k = [\boldsymbol{1} , \widetilde{X}_{k}] G_k$ for all $k$.
        \State Set $\check{X}$, $Z$ and $Y$.
        \State Compute function $\check{q}$ and $\check{h}$ using $\check{X}$, $Z$ and $Y$.
        \State Compute residuals $\hat{\boldsymbol{\eta}}$ and $\hat{\boldsymbol{\zeta}}$ using $\check{q}$, $\check{h}$, $\check{X}$, $Z$ and $Y$.
        \State Obtain $\check{\boldsymbol{\gamma}}$ and $\text{Var}(\check{\boldsymbol{\gamma}})$ as the least square solution of (4.1.2) (in our paper).
        \State Return $R_k^{\text{Point}}$ and $R_k^{\text{Var}}$ to all user $k$.
        \vspace{-.3\baselineskip}

        \Statex \hrulefill
        \vspace{-.3\baselineskip}
        \Statex \textit{user-$k$-side}
        \vspace{-.5\baselineskip}
        \Statex \hrulefill
        \vspace{-.1\baselineskip}
        \State Set $R_k^{\text{Point}}$ and $R_k^{\text{Var}}$.
        \State Compute $\boldsymbol{\beta}_k$, $\text{Var}(\boldsymbol{\gamma}_k)$ and $\text{Var}(\alpha_k)$.
        
        \end{algorithmic}
\end{algorithm}

\subsubsection{Construction of collaborative representations}
\label{sec:method_method_construction}

In the first stage, the construction of collaborative representations, we use the aggregation method proposed in Imakura and Sakurai \citep{imakura2020data}.
First, users generate and share an anchor dataset $X^\text{anc} \in \mathbb{R}^{r \times m}$, which is a shareable dataset consisting of public or randomly constructed dummy data, where $r$ is the number of subjects in the anchor data.
The anchor dataset is used to compute the collaborative representation transformation matrices (described below as $G_k$) required to construct the collaborative representations.
Second, each user $k$ constructs intermediate representations using a linear row-wise dimensionality reduction function $f_k$ such as principal component analysis (PCA) \citep{pearson1901liii}, locality preserving projection (LPP) \citep{he2003locality} and factor analysis (FA) \citep{spearman1904general}.
$f_k$ consists of an intermediate representation transformation matrix $F_{k} \in \mathbb{R}^{m \times \widetilde{m}_k}$ and a shift vector $\mu_k \in \mathbb{R}^{m}$.
$\widetilde{m}_k ~ (< m)$ is the reduced dimension of user $k$.
$\mu_k$ is a mean vector used in $f_k$, taking a non-zero value for centering methods such as PCA and FA, and being zero for methods without centering such as LPP.
Note that $f_k$ is a private function of user $k$ and should not be shared with other users.
As will be described in Section \ref{sec:method_conf_prsv}, the requirement that $f_k$ be a dimensionality reduction function and not shared with other users is important to preserve the confidentiality of the covariates.
The dimensionality reduction function is applied as
\begin{align*}
    \widetilde{X}_{k} &= f_k(X_{k}) = (X_{k} - \boldsymbol{1} \boldsymbol{\mu}_k^\top) F_k \in \mathbb{R}^{n_k \times \widetilde{m}_k}, \\
    \widetilde{X}_{k}^\text{anc} &= f_k(X^\text{anc}) = (X^\text{anc} - \boldsymbol{1} \boldsymbol{\mu}_k^\top) F_k \in \mathbb{R}^{r \times \widetilde{m}_k},
\end{align*}
where $\boldsymbol{1} = [1,\cdots,1]^\top$. 
Moreover, $f_{k}$ is private and can differ from other users.
To consider the constant term in the linear CATE model, we combine $\boldsymbol{1}$ with $\widetilde{X}_{k}$ and $\widetilde{X}_{k}^\text{anc}$.
Formally, we can represent them as
\begin{align*}
    [\boldsymbol{1} , \widetilde{X}_{k}] &=   [\boldsymbol{1} , X_{k} - \boldsymbol{1} \boldsymbol{\mu}_k^\top] \bar{F_k}, \\
    [\boldsymbol{1} , \widetilde{X}_{k}^\text{anc}] &=   [\boldsymbol{1} , X^\text{anc} - \boldsymbol{1} \boldsymbol{\mu}_k^\top] \bar{F_k},
\end{align*}
where
\begin{equation*}
    \bar{F_k} = \left[
                \begin{array}{cc}
                1 & \\ 
                & F_k 
                \end{array}
                \right].
    \label{eq:fbarfunction}
\end{equation*}
Finally, each user $k$ shares the intermediate representations $[\boldsymbol{1} , \widetilde{X}_{k}]$ and $[\boldsymbol{1} , \widetilde{X}_{k}^\text{anc}]$, treatments $Z_{k}$ and outcomes $Y_{k}$ with the analyst.

Next, the analyst transforms the shared intermediate representations into a common form called collaborative representations.
To do this, the analyst computes the transformation functions for all $k$ such that
\begin{equation*}
    [\boldsymbol{1} , \widetilde{X}_{k}^\text{anc}] G_k \approx [\boldsymbol{1} , \widetilde{X}_{k'}^\text{anc}] G_{k'} \in \mathbb{R}^{r \times \check{m}} ~ (k \neq k'),
    \label{eq:g_cond}
\end{equation*}
where $G_k \in \mathbb{R}^{(\widetilde{m}_k + 1) \times \check{m}}$ and $\check{m}$ is the dimension of collaborative representations.
Then, we have the collaborative representation for $k$ as
\begin{equation*}
    \check{X}_k = [\boldsymbol{1} , \widetilde{X}_{k}] G_k \in \mathbb{R}^{n_k \times \check{m}}.
\end{equation*}
Let
\begin{equation*}
    \left[ [\boldsymbol{1} , \widetilde{X}_{1}^\text{anc}] , \cdots , [\boldsymbol{1} , \widetilde{X}_{c}^\text{anc}] \right] =
            [U_1 , U_2]
            \left[
                \begin{array}{cc}
                \Sigma_1 & \\ 
                 & \Sigma_2
                \end{array}
            \right]
            \left[
                \begin{array}{c}
                V_1^\top\\ 
                V_2^\top
                \end{array}
            \right]
        \approx U_1 \Sigma_1 V_1^\top
\end{equation*}
be a low-rank approximation based on the singular value decomposition of the matrix combining $[\boldsymbol{1} , \widetilde{X}_{k}^\text{anc}]$,
where $\Sigma_1 \in \mathbb{R}^{\check{m} \times \check{m}}$ is a diagonal matrix whose diagonal entries are the top $\check{m}$ singular values in descending order,
and $U_1$ and $V_1$ are column orthogonal matrices whose columns are the corresponding left and right singular vectors, respectively.
In this study, we set $\check{m} = m$ for simplicity; see Section \ref{sec:simulation} for discussion of this choice.
The matrix $G_k$ is then computed as
\begin{equation*}
    G_k = \left( [\boldsymbol{1} , \widetilde{X}_{k}^\text{anc}] \right)^\dagger U_1,
\end{equation*}
where $\dagger$ denotes the Moore--Penrose inverse.
Then, the collaborative representations are given by
\begin{equation*}
    \check{X} =   \left[
                                \begin{array}{c}
                                \check{X}_1\\ 
                                \vdots \\
                                \check{X}_c
                                \end{array}
                            \right]
    \in \mathbb{R}^{n \times \check{m}}.
\end{equation*}

\subsubsection{Estimation by the analyst}
\label{sec:method_method_esta}
In the second stage, the estimation by the analyst, the analyst estimates their own CATE model based on the DML procedure that we described in Section \ref{sec:linear_DML}.
The analyst considers the data-generation process as
\begin{align*}
    y_i &= \theta(\check{\boldsymbol{x}}_i) \cdot z_i + u(\check{\boldsymbol{x}}_i) + \varepsilon_i, \\
    z_i &= h(\check{\boldsymbol{x}}_i) + \eta_i,
\end{align*}
and the linear CATE model as
\begin{equation*}
    \theta(\check{\boldsymbol{x}}_i) = \check{\boldsymbol{x}}_i^\top \boldsymbol{\gamma} = \check{x}_i^1 \gamma^1 + \cdots + \check{x}_i^{\check{m}} \gamma^{\check{m}},
\end{equation*}
where $\boldsymbol{\gamma} = [\gamma^1, \cdots, \gamma^{\check{m}} ]^\top \in \mathbb{R}^{\check{m}}$ and $\check{\boldsymbol{x}}_i$ is the collaborative representation of $\boldsymbol{x}_i$.
The score vectors for the analyst are defined as
\begin{align*}
    \check{\psi}(\check{\boldsymbol{x}}_i; \check{\boldsymbol{\gamma}}, \check{q}, \check{h}) &= \check{\boldsymbol{x}}_i (z_i - \check{h}(\check{\boldsymbol{x}}_i))(y_i - \check{q}(\check{\boldsymbol{x}}_i) - \check{\boldsymbol{x}}_i^\top \check{\boldsymbol{\gamma}} (z_i - \check{h}(\check{\boldsymbol{x}}_i))), \\
    \check{\psi}^a (\check{\boldsymbol{x}}_i; \check{h}) &= (z_i - \check{h}(\check{\boldsymbol{x}}_i))^2 \check{\boldsymbol{x}}_i \check{\boldsymbol{x}}_i^\top.
\end{align*}
where $\check{q}$ and $\check{h}$ are the analyst's functions, and $\check{\boldsymbol{\gamma}} \in \mathbb{R}^{\check{m}}$ is the analyst's estimates.

The analyst's purpose in the second stage is to obtain $\check{\boldsymbol{\gamma}}$.
To accomplish this, first, the analyst constructs $\check{q}$ and $\check{h}$ from $\check{X}$, $Z$ and $Y$ using machine learning models to estimate residuals $\zeta_i$ and $\eta_i$.
The estimated residuals are expressed as
\begin{align*}
    \hat{\zeta}_i &= y_i - \check{q}(\check{\boldsymbol{x}}_i), \\
    \hat{\eta}_i &= z_i - \check{h}(\check{\boldsymbol{x}}_i).
\end{align*}
Second, the analyst calculates $\check{\boldsymbol{\gamma}}$ and $\text{Var}(\check{\boldsymbol{\gamma}}) \in \mathbb{R}^{\check{m} \times \check{m}}$.
$\check{\boldsymbol{\gamma}}$ is the least squares solution of the following regression equation:
\begin{equation*}
    \hat{\zeta_i} = \hat{\eta}_i \check{\boldsymbol{x}}_i^\top \boldsymbol{\gamma} + \varepsilon_i.
    \label{eq:analyst_beta}
\end{equation*}
That is,
\begin{equation*}
    \check{\boldsymbol{\gamma}} = \argmin_{\boldsymbol{\gamma}'} \left( \sum_{i} \left( \hat{\zeta}_i - \hat{\eta}_i \check{\boldsymbol{x}}_i^\top \boldsymbol{\gamma}' \right)^2 \right).
\end{equation*}
The analyst can calculate the variance of $\check{\boldsymbol{\gamma}}$ in (\ref{eq:sigma_hat_2}), (\ref{eq:J_hat}) and (\ref{eq:beta_var}), replacing $\psi$, $\psi^a$ and $\hat{\boldsymbol{\beta}}$ with $\check{\psi}$, $\check{\psi}^a$ and $\check{\boldsymbol{\gamma}}$, respectively.

\subsubsection{Estimation by the users}
\label{sec:method_method_estu}
In the final stage, the estimation by the users, the users estimate their own CATE model based on parameters that the analyst returns.
The analysts's CATE model is written as
\begin{align*}
    \check{X}_k \check{\boldsymbol{\gamma}} &= [\boldsymbol{1} , \widetilde{X}_{k}] G_k \check{\boldsymbol{\gamma}} \\
                        &= [\boldsymbol{1} , X_{k} - \boldsymbol{1} \boldsymbol{\mu}_k^\top] \bar{F_k} G_k \check{\boldsymbol{\gamma}}.
\end{align*}
Then, for user $k$, subject $i$'s estimated CATE $\hat{\tau}_i$ is written as
\begin{equation}
    \begin{aligned}
        \hat{\tau}_i &= \gamma_k^0 + (x_i^1 - \mu_k^1)\gamma_k^1 + \cdots + (x_i^m - \mu_k^m)\gamma_k^m \\
                        &= [1, -\mu_k^1, \cdots, -\mu_k^m] \boldsymbol{\gamma}_k + \gamma_k^1 x_i^1 + \cdots + \gamma_k^m x_i^m \\
                        &= \alpha_k + \gamma_k^1 x_i^1 + \cdots + \gamma_k^m x_i^m
    \end{aligned}
    \label{eq:gamma_cate}
\end{equation}
where
\begin{equation*}
    \boldsymbol{\gamma}_k = [\gamma_k^0,\gamma_k^1,\cdots,\gamma_k^m]^\top = \bar{F_k} G_k \check{\boldsymbol{\gamma}},
\end{equation*}
and
\begin{equation}
    \alpha_k = [1, -\mu_k^1, \cdots, -\mu_k^m] \boldsymbol{\gamma}_k.
    \label{eq:alpha}
\end{equation}
For user $k$, $\alpha_k$ is a constant term and $\gamma_k^1,\cdots,\gamma_k^m$ are coefficients for $x_i^1,\cdots,x_i^m$ in their CATE model.
Thus, user $k$'s coefficients are expressed as $\boldsymbol{\beta}_k = [ \alpha_k, \gamma_k^1, \cdots, \gamma_k^m ]^\top$.

The variance of $\boldsymbol{\gamma}_k$ is
\begin{equation*}
    \text{Var}(\boldsymbol{\gamma}_k) = \bar{F_k} G_k \text{Var}(\check{\boldsymbol{\gamma}}) G_k^\top \bar{F_k}^\top.
\end{equation*}
Then, the variance of $\alpha_k$ is
\begin{equation}
    \text{Var}(\alpha_k) = [1, -\mu_k^1, \cdots, -\mu_k^m] \text{Var}(\boldsymbol{\gamma}_k) 
                            \left[
                                \begin{array}{c}
                                1\\
                                -\mu_k^1 \\
                                \vdots \\
                                -\mu_k^m
                                \end{array}
                            \right].
    \label{eq:var_alpha}
\end{equation}

User $k$ can statistically test their CATE model if they obtains $\boldsymbol{\beta}_k$, $\text{Var}(\boldsymbol{\gamma}_k)$ and $\text{Var}(\alpha_k)$.
To accomplish this, the following procedure is conducted in the final stage.
First, the analyst returns $R_k^{\text{Point}} = G_k \check{\boldsymbol{\gamma}}$ and $R_k^{\text{Var}} = G_k \text{Var}(\check{\boldsymbol{\gamma}}) G_k^\top$ to user $k$.
Then, user $k$ calculates $\boldsymbol{\gamma}_k$ and $\text{Var}(\boldsymbol{\gamma}_k)$ as
\begin{align*}
    \boldsymbol{\gamma}_k &=  \bar{F_k} R_k^{\text{Point}}, \\
    \text{Var}(\boldsymbol{\gamma}_k) &= \bar{F_k} R_k^{\text{Var}} \bar{F_k}^\top.
\end{align*}
Moreover, user $k$ calculates $\alpha_k$ and $\text{Var}(\alpha_k)$ using (\ref{eq:alpha}) and (\ref{eq:var_alpha}), respectively.
Then, user $k$ obtains $\boldsymbol{\beta}_k$, $\text{Var}(\boldsymbol{\gamma}_k)$ and $\text{Var}(\alpha_k)$, and can test statistically $\boldsymbol{\beta}_k$.

For user $k$, subject $i$'s estimated CATE $\hat{\tau}_i$ are calculated from (\ref{eq:gamma_cate}).
To derive the variance of subject $i$'s CATE, rewrite (\ref{eq:gamma_cate}) as
\begin{equation*}
    \hat{\tau}_i = [1, x_i^1 - \mu_k^1, \cdots, x_i^m - \mu_k^m] \boldsymbol{\gamma}_k
\end{equation*}
Then,
\begin{equation*}
    \text{Var}(\hat{\tau}_i) = [1, x_i^1 - \mu_k^1, \cdots, x_i^m - \mu_k^m] \text{Var}(\boldsymbol{\gamma}_k)
                            \left[
                                \begin{array}{c}
                                1\\
                                x_i^1 - \mu_k^1 \\
                                \vdots \\
                                x_i^m - \mu_k^m
                                \end{array}
                            \right].
\end{equation*}
Thus, user $k$ can test $\hat{\tau}_i$ statistically based on $\text{Var}(\hat{\tau}_i)$.

\subsection{Correctness of DC-DML}
\label{sec:method_correctness}
In this section, using a similar approach to Imakura et al. \citep{imakura2023dccox}, we discuss the correctness of DC-DML.
We now consider the following subspace for each party $k$:
\begin{equation*}
    \mathcal{S}_k = \mathcal{R}(\bar{F_k} G_k) \subset \mathbb{R}^{m+1}, ~ \text{dim}(\mathcal{S}_k) = \check{m},
\end{equation*}
where $\mathcal{R}$ denotes the range of a matrix.
We define centralized analysis (CA) as the collection and analysis of the raw data, i.e., $X$, $Z$ and $Y$, in one place.
While this represents the ideal scenario, it is infeasible when raw data contain confidential information.
Then, we have Theorem \ref{thm:correctness}, whose proof is given in \ref{app:proof_correctness}.
\begin{thm}
    Let $\boldsymbol{\beta}_{\textup{CA}}$ be the coefficients, and $q_{\textup{CA}}$ and $h_{\textup{CA}}$ be the functions computed in CA.
    Moreover, consider $\boldsymbol{\gamma}_{\textup{CA}} = [\gamma_{\textup{CA}}^0, \gamma_{\textup{CA}}^1, \cdots, \gamma_{\textup{CA}}^m]^\top$ such that
    \begin{equation*}
        \gamma_{\textup{CA}}^0 = [1, \mu_1^1, \cdots, \mu_1^m] \boldsymbol{\beta}_{\textup{CA}}, \gamma_{\textup{CA}}^j = \beta_{\textup{CA}}^j ~ (j \neq 0).
    \end{equation*}
    If $\boldsymbol{\gamma}_{\textup{CA}} \in \mathcal{S}_1 = \cdots = \mathcal{S}_c$,
    $\boldsymbol{\mu}_1 = \cdots = \boldsymbol{\mu}_c$,
    $\textup{rank}(X^{\textup{anc}} \bar{F}_k) = \widetilde{m}$ for all $k \in \{1,\cdots, c \}$,
    $q_{\textup{CA}}(\boldsymbol{x}_i) = \check{q}(\check{\boldsymbol{x}}_i)$ and
    $h_{\textup{CA}}(\boldsymbol{x}_i) = \check{h}(\check{\boldsymbol{x}}_i)$ for all $i \in \{1,\cdots, n \}$,
    then $\boldsymbol{\beta}_{\textup{CA}} = \boldsymbol{\beta}_k$.
    \label{thm:correctness}
\end{thm}
Theorem \ref{thm:correctness} implies that, under the stated assumptions, DC-DML can recover the same parameters as those obtained in CA.
Although it is unlikely that all the assumptions are satisfied, Theorem \ref{thm:correctness} provides some hints for the dimensionality reduction methods as we describe in Section \ref{sec:method_dimmethod}.

\subsection{A dimension reduction method for DC-DML}
\label{sec:method_dimmethod}
In this section, we propose a dimensionality reduction method for DC-DML with reference to Theorem \ref{thm:correctness}.
The choice of dimensionality reduction method critically affects the performance of DC-DML.
In DC-DML, the coefficients $\boldsymbol{\gamma}_k$ are obtained from the subspace $\mathcal{S}_k$.
If we could obtain $\boldsymbol{\gamma}_k$ close to $\boldsymbol{\gamma}_{\text{CA}}$, we could achieve $\boldsymbol{\beta}_k$ close to $\boldsymbol{\beta}_{\text{CA}}$.
To accomplish this, the subspace $\mathcal{S}_k$ should contain $\boldsymbol{\gamma}_{\text{CA}}$.

Using an approach similar to Imakura et al. \citep{imakura2023dccox}, we consider a dimensionality reduction method for DC-DML.
Algorithm \ref{alg:bstrap} is the pseudo-code of that.
The basic idea of the approach is that each user $k$ constructs $F_k$ based on estimates of $\boldsymbol{\gamma}_{\text{CA}}$ obtained from their own local dataset.
The estimate of $\boldsymbol{\gamma}_{\text{CA}}$ obtained by each user could be a good approximation of $\boldsymbol{\gamma}_{\text{CA}}$.
Thus, a bootstrap-based method can be a reasonable choice to achieve a good subspace.
Let $0 < p < 1$ be a parameter for the sampling rate.
In the bootstrap-based method, first, through the DML procedure by random samplings of size $p n_k$, user $k$ obtains estimates of $\boldsymbol{\gamma}_{\text{CA}}$ as $\boldsymbol{\gamma}_1, \cdots, \boldsymbol{\gamma}_{\widetilde{m}_k}$.
User $k$ then constructs the matrix $F_k = [\bar{\boldsymbol{\gamma}}_1 , \cdots , \bar{\boldsymbol{\gamma}}_{\widetilde{m}_k} ]$, where $\bar{\boldsymbol{\gamma}}_b \in \mathbb{R}^{m}$ is a vector that excludes the first element of $\boldsymbol{\gamma}_b \in \mathbb{R}^{m+1}$.
The reason for excluding the first element of $\boldsymbol{\gamma}_b$ is that $F_k$ should be an intermediate representation transformation matrix for $m$-dimensional covariates.
However, since the first column of $\bar{F}_k$ is $[1,0,\cdots,0]^\top$ as in (\ref{eq:fbarfunction}), the subspace contains the full range that the first element of $\boldsymbol{\gamma}_{\text{CA}}$ can take.

\begin{algorithm}[tb]
    \caption{A bootstrap-based dimensionality reduction for DC-DML}
    \label{alg:bstrap}
    \begin{algorithmic}[1]
        \Statex \textbf{Input: covariates $X_k$, treatments $Z_k$, outcomes $Y_k$}.
        \Statex \textbf{Output: $F_k$}.
        \vspace{-.3\baselineskip}
        
        \Statex \hrulefill
        \State \textbf{for} $b = 1, \cdots, \widetilde{m}_k$ : \textbf{do}
        \State \qquad Set $X_k^b$, $Z_k^b$ and $Y_k^b$ by a random sampling of size $p n_k$ from $X_k$, $Z_k$ and $Y_k$.
        \State \qquad Compute $\boldsymbol{\gamma}_b$ from $X_k^b$, $Z_k^b$ and $Y_k^b$ by the DML procedure.
        \State \textbf{end for}
        \State Return $F_k = [\bar{\boldsymbol{\gamma}}_1 , \cdots , \bar{\boldsymbol{\gamma}}_{\widetilde{m}_k} ]$.
        
        \end{algorithmic}
\end{algorithm}

Note that while the proposed bootstrap-based method is designed to construct a subspace $\mathcal{S}_k$ that likely contains $\boldsymbol{\gamma}_{\text{CA}}$, this does not imply that all assumptions of Theorem \ref{thm:correctness} are strictly satisfied.
In practice, it is difficult to ensure the exact validity of all assumptions in distributed settings.
Therefore, Algorithm \ref{alg:bstrap} should be regarded as a heuristic approach that empirically yields 
values $\boldsymbol{\beta}_k$ close to $\boldsymbol{\beta}_{\text{CA}}$, rather than a method that provides theoretical guarantees for its accuracy.
Theoretical analysis of the approximation error when the assumptions are not fully satisfied is left as an important direction for future work.
Similar error evaluations have been performed in related settings, such as \cite{imakura2021accuracy}, and we anticipate that a similar analysis could be applicable in our context in future work.

In addition, as with Imakura et al. \citep{imakura2023dccox}, the subspace can be constructed by combining multiple dimensionality reduction methods.
Let $F_{\text{BS}} \in \mathbb{R}^{m \times \widetilde{m}_{\text{BS}}}$ and $F_{\text{DR}} \in \mathbb{R}^{m \times \widetilde{m}_{\text{DR}}}$ be the intermediate representation transformation matrices of the bootstrap-based and other methods, respectively, where $\widetilde{m}_k = \widetilde{m}_{\text{BS}} + \widetilde{m}_{\text{DR}}$.
Then $F_k$ is
\begin{equation*}
    F_k = [F_{\text{BS}} , F_{\text{DR}}].
\end{equation*}

\subsection{Discussion on confidentiality preservation}
\label{sec:method_conf_prsv}

In this section, we discuss how DC-DML preserves the confidentiality of covariates.
DC-DML prevents the original covariates from being recovered or inferred from the shared intermediate representations by the following two layers of confidentiality, inherited from DC technology \citep{imakura2020data}.

The first layer is to construct intermediate representations by dimensionality reduction of the covariates.
Since the dimensionality reduction results in the loss of some of the original covariate information, it becomes infeasible to reconstruct the original covariates from the intermediate representations.

The second layer is that each user $k$ does not share the dimensionality reduction function $f_k$ with the other users.
If $f_k$ were to be shared, other parties could potentially construct an approximate inverse function $f_k^{-1}$, and then infer the original covariates using $f_k^{-1}$ for the shared intermediate representations of user $k$.
The construction of $f_k^{-1}$ is possible for those who know both $X^\text{anc}$ and $\widetilde{X}_{k}^\text{anc}$, but as long as the DC protocol is properly followed, only user $k$ has access to both $X^\text{anc}$ and $\widetilde{X}_{k}^\text{anc}$, thus preventing reconstruction..

Various attacks are possible to recover or infer the original covariates from the shared intermediate representations.
For example, it is a collusion between users and analysts.
Depending on the way of collusion, the confidentiality of the covariates may be violated, as discussed by \cite{imakura2021accuracy}, for more details, see.
However, we believe that future research could create robust DC technology against a variety of attacks.

\subsection{Advantage and disadvantage}
\label{sec:method_ad_disad}

In this section, we describe the advantages and disadvantages of DC-DML.
Those are attributed to DC-QE or DML with the linear CATE model described in Section \ref{sec:linear_DML}.

There are four advantages.
First, as mentioned in Section \ref{sec:method_conf_prsv}, the dimensionality reduction preserves the confidentiality of the covariates.
Second, since DC-DML enables the collection of a larger number of subjects from distributed data, and accumulate a knowledge base, it is possible to obtain better performance than in individual analysis.
Third, DC-DML does not require repetitive communication (communication-efficient).
Finally, DC-DML is robust against model mis-specification due to its semi-parametric nature

Conversely, there are three disadvantages.
First, DC-DML cannot be used directly if confidential information is included in the treatments or outcomes.
Solutions to this issue include, for example, transforming treatments or outcomes into a format that preserves confidentiality through encryption, etc.
Second, the performance of the resulting model may deteriorate because the dimensionality reduction reduces the information in the raw data.
However, obtaining more subjects from the distributed data can improve performance more than the deterioration caused by dimensionality reduction.
Finally, DC-DML may yield misleading conclusions if the assumption of linearity in the CATE model does not hold.

\section{Simulations}
\label{sec:simulation}

In this section, we describe three simulations we conducted to demonstrate the effectiveness of DC-DML.
\footnote{We performed the simulations using Python codes, which are available from the corresponding author by reasonable request.}
In Simulation I, which used synthetic data, we showed that DC-DML was useful in the setting of data distributions that would be incorrectly estimated by individual analysis (IA), the analysis performed by the user using only their own dataset.
In Simulation II, which used semi-synthetic data generated from the infant dataset, we compared the performances between DC-DML and other existing methods.
In Simulation III, which used real-world data from financial assets and jobs datasets, we investigated the robustness of DC-DML performance in the presence of heterogeneous numbers of subjects or proportions of treatment group per party.

In Simulation II, the methods compared to DC-DML were as follows.
First are the DMLs in IA and CA, that we denote for convenience as IA-DML and CA-DML, respectively.
As described above, IA-DML and CA-DML were targets that DC-DML was intended to perform better and closer to, respectively.
Second, GRF and X-Learner in IA and CA, denoted for convenience with the prefixes IA- and CA-: IA-GRF, CA-GRF, IA-XL, and CA-XL.
Finally, the distributed data analysis methods SR, FedCI, CausalRFF, Federated IPW-ML (FIM) and DC-QE.

In the simulations, we set up the machine learning models used in DML and X-Learner as follows.
First, both machine learning models used to estimate $q$ and $h$ in the DML methods (IA-DML, CA-DML and DC-DML) in Simulation I were random forests.
Second, the settings of the machine learning models used in Simulations II and III are as follows.
We selected the machine learning models used to estimate $q$ in the DML methods based on our preliminary simulations from among five regression methods: multiple regression, random forests, K-nearest neighbor, light gradient boosting machine (LGBM) and support vector machine (SVM) with standardization.
In the preliminary simulations for $q$, we repeated the regression of the outcomes $Y$ on the covariates $X$ for the candidate methods in the CA setting 50 times.
We evaluated their predictive performance using the root mean squared error (RMSE) and selected the method with the smallest average RMSE.
For $h$, similarly, we selected by preliminary simulations among five classification methods: logistic regression, random forests, K-nearest neighbor, LGBM and SVR.
In our preliminary simulations for $h$, we followed a similar procedure, where we evaluated the predicted probabilities of the treatments $Z$ using the Brier score, and selected the method with the smallest average Brier score.
Here, Brier Score is a mean squared error measure of probability prediction \citep{brier1950verification}.
The hyperparameters for the candidate methods were the default values shown in \ref{app:hparams_cand}.
As a result of our preliminary simulations, we chose SVM as the machine learning models to be used in the estimation of $q$ and $h$, both in Simulation II.
For the financial data set in Simulation III, we set $q$ and $h$ to multiple regression and logistic regression, respectively.
For the jobs dataset, we set $q$ and $h$ to multiple regression and random forest, respectively.
We used the same models for the outcome and propensity score in X-Learner (IA-XL and CA-XL) as in DML in all simulations.
The results of the preliminary simulations are in \ref{app:pre_exp}.

The settings used in DC-DML were as follows.
As the intermediate representation construction methods, we used six: PCA, LPP, FA, PCA+B, LPP+B, and FA+B. Here, +B represents the combination with the bootstrap-based dimensionality reduction method.
The intermediate representation dimension and collaborative representation dimension were set as $\widetilde{m}_k = m-1$ for all $k$ and $\check{m} = m$, respectively.
In combination with the bootstrap-based dimensionality reduction method, the BS dimension was set to three in Simulation I and $\lceil 0.1m \rceil$ in Simulation II and III.
Here,$\lceil \cdot \rceil$ is the ceiling function.
Anchor data were generated from uniform random numbers of maximum and minimum values for each covariate and their subject number $r$ was the same as the original dataset $X$.

While we fixed the dimension of collaborative representations $\check{m}$ to $m$ in all simulations, following \cite{kawamata2024collaborative}, for a consistent and simple baseline, we have not investigated whether this choice is optimal in our setting. Nonetheless, as shown later in this section, our method demonstrated consistently good performance under this setting. The related study \citep{imakura2021interpretable} implied that the performance of the DC framework can be sensitive to this parameter. Therefore, investigating data-driven strategies for selecting $\check{m}$ is an important direction for future work.

As described below, in Simulation II, we used the eight methods described in Section \ref{sec:relatedwork} as comparison methods: DML, GRF, X-Learner, SR, FedCI, CausalRFF, FIM and DC-QE.
Also, in Simulation III, we used DML and SR as comparison methods.
See \ref{app:methods_settings} for the detailed settings of those methods.

In Simulations II and III, the evaluation measures were how close the estimates were to the benchmarks.
The benchmarks in Simulations II and III were the true values and the average values of 50 CA-DML trials, respectively.
In this study, we considered five evaluation measures: the RMSE of CATE, the consistency rate of significance of CATE, the RMSE of coefficients, the consistency rate of significance of coefficients, and the ATE.
To enable comparison with existing ATE estimation methods for distributed data, i.e. FIM and DC-QE, we also reported ATE values as an auxiliary evaluation metric, computed by averaging estimated CATEs over the population.
Note that, for simplicity of explanation, results other than the RMSE of CATE in Simulation III were shown in \ref{app:expIII}.

\subsection*{The RMSE of CATE}
The RMSE of CATE is a measure of how much the predicted CATE matched the benchmark CATE as
\begin{equation*}
    \sqrt{\frac{1}{N} \Sigma_i \left( \tau_{i,\textup{BM}} - \hat{\tau}_i \right)^2},
\end{equation*}
where $\tau_{i,\textup{BM}}$ is the benchmark CATE and $\hat{\tau}_i$ is the estimated CATE.
The lower this measure, the better.
Since FIM and DC-QE are not models for CATE estimation, this measure was not calculated in those models.

\subsection*{The consistency rate of significance of CATE}
The consistency rate of significance of CATE is the rate where the estimated CATE obtained a statistical test result of ``significantly positive,'' ``significantly negative,'' or ``not significant'' in 5\% level when the benchmark CATE is positive, negative, or zero, respectively.
For Simulation III, we set the benchmark CATE is positive, negative, or zero when the estimate in CA-DML is significantly positive, significantly negative or not significant, respectively, at 5\% level.
The higher this measure, the better.
This measure was not calculated in X-Learner, FedCI, CausalRFF, FIM and DC-QE, because statistical testing methods for CATE are not provided in those methods.

\subsection*{The RMSE of coefficients}
The RMSE of coefficients is a measure of how much the estimated coefficients matched that of the benchmark as
\begin{equation*}
    \sqrt{\frac{1}{m+1} \Sigma_{j} \left( \beta_{\textup{BM}}^j - \beta_k^j \right)^2},
\end{equation*}
where $\beta_{BM}^j$ is the benchmark coefficient.
The lower this measure, the better.
Since GRF, X-Learner, FedCI, CausalRFF, FIM and DC-QE are not models for linear CATE estimation, this measure was not calculated in those models.

\subsection*{The consistency rate of significance of coefficients}
The consistency rate of significance of coefficients is the rate where the estimated coefficient obtained a statistical test result of ``significantly positive,'' ``significantly negative,'' or ``not significant'' at 5\% level when the benchmark coefficient is positive, negative, or zero, respectively.
For Simulation III, we set the benchmark coefficient is positive, negative, or zero when the estimate in CA-DML is significantly positive, significantly negative or not significant, respectively, at 5\% level.
The higher this measure, the better.
As with the RMSE of coefficients, this measure was not calculated in GRF, X-Learner, FedCI, CausalRFF, FIM and DC-QE.

\subsection*{The average treatment effect}
The ATE is a measure of the overall causal effect of the treatments on the population. In this study, we compute the ATE by averaging the estimated CATEs as
\begin{equation*}
    \frac{1}{N} \Sigma_i (\hat{\tau}_i),
\end{equation*}
The closer this estimate is to the true ATE, the better.

Note that our evaluations in the simulations were conducted entirely in-sample.
Since our primary objective is to estimate the CATE, rather than to predict individual outcomes or treatments, we did not perform standard out-of-sample validation using held-out test sets.
This evaluation strategy follows common practice in causal inference studies using DML (e.g., \cite{baiardi2024value}), where the emphasis lies on in-sample estimation validity.

\subsection{Simulation I: Proof of concepts in synthetic data}
\label{sec:simulation-poc}
\begin{figure}[tb]
  \centering
  \includegraphics[width=0.3\linewidth]{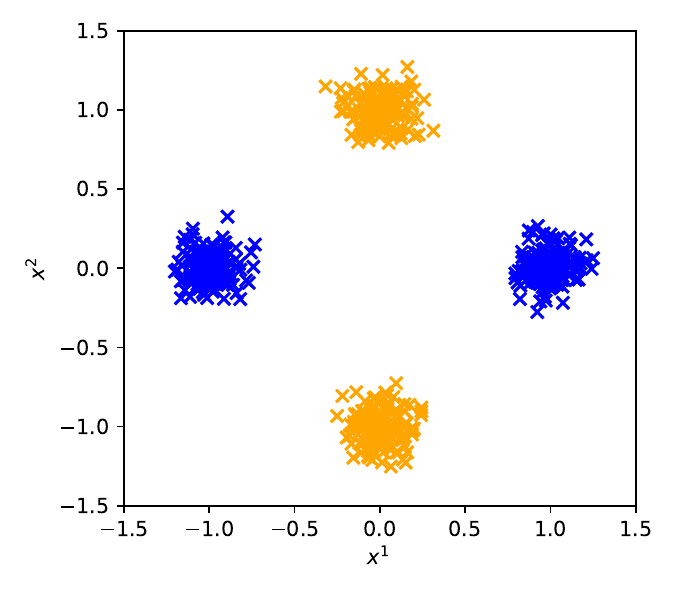}
  \caption{The data distribution in Simulation I. The blue and orange markers represent the covariates of party 1 and 2, respectively.}
  \label{fig:exp1_datadist}
\end{figure}

In Simulation I, we considered the situation where two parties ($c=2$) each had a dataset consisting of 300 subjects ($n_k=300$) and 10 covariates ($m=10$).
The covariates $x_i^1$ and $x_i^2$ were distributed as in Fig. \ref{fig:exp1_datadist}, where blue and orange markers represented those of party 1 and 2, respectively.
The other covariates were drawn from a normal distribution $\mathcal{N}(0,1)$.

The setting of the data-generation process in Simulation I was
\begin{align*}
    \theta(\boldsymbol{x}_i) &= 1 + x_i^1 + x_i^2, \\
    u(\boldsymbol{x}_i) &= |x_i^1| + |x_i^2|, \\
    h(\boldsymbol{x}_i) &= \frac{1}{1-\exp{(-x_i^1 - x_i^2)}}, \\
    \varepsilon_i &\sim \mathcal{N}(0,0.1) , \\
    \eta_i &=   \left\{
                    \begin{array}{ll}
                    1 - h(\boldsymbol{x}_i) & \textup{with probability} ~ h(\boldsymbol{x}_i) \\
                    - h(\boldsymbol{x}_i) & \textup{with probability} ~ 1 - h(\boldsymbol{x}_i)
                    \end{array}
                \right. .
\end{align*}
In this setting, the constant term and the coefficients of $x_i^1$ and $x_i^2$ should have been estimated to be significantly one.

Party 1 had the dataset with the broad distribution for covariate 1 and the narrow distribution for covariate 2.
Thus, party 1 would be able to estimate the coefficient of covariate 1 relatively correctly in IA-DML, but would have difficulty estimating the coefficient of covariate 2.
The opposite was true for party 2.

The result of Simulation I is shown in Table \ref{tab:exp1}.
In CA-DML, the constant term and the coefficients of $x_i^1$ and $x_i^2$ were significant and close to their true values, and the coefficients of the other covariates are not significant.
Thus, CA-DML obtained reasonable estimation results.
Conversely, IA-DML obtained some wrong estimation results in both parties.

\begin{table}[tb]
\centering
    \begin{tabular}{lrlrlrlrlrl}
    \hline \hline
    \multicolumn{1}{l|}{\multirow{2}{*}{Covariates}} & \multicolumn{2}{l|}{\multirow{2}{*}{CA-DML}} & \multicolumn{4}{l|}{IA-DML} & \multicolumn{4}{l}{DC-DML(PCA+B)} \\ \cline{4-11} 
    \multicolumn{1}{l|}{} & \multicolumn{2}{l|}{} & \multicolumn{2}{l}{Party 1} & \multicolumn{2}{l|}{Party 2} & \multicolumn{2}{l}{Party 1} & \multicolumn{2}{l}{Party 2} \\ \hline
    \multicolumn{1}{l|}{const.} & 1.0283 & \multicolumn{1}{l|}{**} & 0.6300 & ** & 1.0068 & \multicolumn{1}{l|}{**} & 0.9563 & ** & 1.0207 & ** \\
    \multicolumn{1}{l|}{$x^1$} & 1.0610 & \multicolumn{1}{l|}{**} & 0.9952 & ** & 0.6950 & \multicolumn{1}{l|}{} & 1.0578 & ** & 1.0119 & ** \\
    \multicolumn{1}{l|}{$x^2$} & 1.1290 & \multicolumn{1}{l|}{**} & 2.0799 &  & 1.0359 & \multicolumn{1}{l|}{**} & 0.9042 & ** & 0.9761 & ** \\
    \multicolumn{1}{l|}{$x^3$} & -0.1628 & \multicolumn{1}{l|}{} & -0.1161 &  & -0.3211 & \multicolumn{1}{l|}{*} & -0.0961 &  & -0.1269 &  \\
    \multicolumn{1}{l|}{$x^4$} & -0.1595 & \multicolumn{1}{l|}{} & -0.0220 &  & -0.0021 & \multicolumn{1}{l|}{} & -0.1489 &  & -0.1264 &  \\
    \multicolumn{1}{l|}{$x^5$} & 0.0566 & \multicolumn{1}{l|}{} & -0.0924 &  & 0.1679 & \multicolumn{1}{l|}{} & 0.0147 &  & 0.0190 &  \\
    \multicolumn{1}{l|}{$x^6$} & 0.0905 & \multicolumn{1}{l|}{} & 0.3367 & * & -0.2482 & \multicolumn{1}{l|}{} & 0.1608 &  & 0.1354 &  \\
    \multicolumn{1}{l|}{$x^7$} & -0.0434 & \multicolumn{1}{l|}{} & 0.0771 &  & -0.1849 & \multicolumn{1}{l|}{} & -0.0317 &  & -0.0960 &  \\
    \multicolumn{1}{l|}{$x^8$} & -0.0494 & \multicolumn{1}{l|}{} & -0.0397 &  & -0.1370 & \multicolumn{1}{l|}{} & -0.0336 &  & -0.0065 &  \\
    \multicolumn{1}{l|}{$x^9$} & 0.0607 & \multicolumn{1}{l|}{} & -0.0955 &  & 0.1867 & \multicolumn{1}{l|}{} & 0.0418 &  & 0.0536 &  \\
    \multicolumn{1}{l|}{$x^{10}$} & -0.1773 & \multicolumn{1}{l|}{} & -0.2022 &  & -0.0543 & \multicolumn{1}{l|}{} & -0.1270 &  & -0.1391 &  \\ \hline \hline
    \multicolumn{3}{l}{** $p<0.01$, ~ * $p<0.05$} & \multicolumn{1}{l}{} &  & \multicolumn{1}{l}{} &  & \multicolumn{1}{l}{} &  & \multicolumn{1}{l}{} & 
    \end{tabular}
\caption{The result of Simulation I.}
\label{tab:exp1}
\end{table}

DC-DML conducted by the PCA+B dimensionality reduction method obtained estimation results similar to those obtained by CA-DML.
This result suggested that even in situations where it was difficult to make valid estimates using IA-DML, DC-DML could produce more valid estimates than that.

\subsection{Simulation II: Comparison with other methods in semi-synthetic data}
\label{sec:simulation-comp}
In Simulation II, we evaluated the performance of DC-DML on semi-synthetic data generated from the infant health and development program (IHDP) dataset.
The IHDP dataset is ``a randomized simulation that began in 1985, targeted low-birth-weight, premature infants, and provided the treatment group with both intensive high-quality child care and home visits from a trained provider.'' \citep{hill2011bayesian}
\footnote{The IHDP dataset is published as the supplementary file (\url{https://www.tandfonline.com/doi/abs/10.1198/jcgs.2010.08162}).}
The IHDP dataset consisted of 25 covariates and a treatment variable and did not include an outcome variable.
Thus, the outcomes needed to be synthetically set up for the simulation.
Hill \cite{hill2011bayesian} removed data on all children with non-white mothers from the treatment group to imbalance the covariates in the treatment and control groups.
We also followed to this end, and as a result, the dataset in Simulation II consisted of 139 subjects in the treatment group and 608 subjects in the control group.

In Simulation II, we considered the situation where three parties ($c=3$) each had a dataset consisting of 249 subjects ($n_k=249$) and 25 covariates ($m=25$).
The ratios of the treatment groups were equal in all parties.
We set the outcome variable in Simulation II as
\begin{align*}
    \theta(\boldsymbol{x}_i) &= \textup{const.} + \boldsymbol{x}_i^\top \boldsymbol{\beta}, \\
    \boldsymbol{\beta} &= \left[ \frac{1}{\sigma^1}, 0, -\frac{1}{\sigma^3}, \frac{1}{\sigma^4}, 0, -\frac{1}{\sigma^6}, \cdots \right]^\top, \\
    \textup{const.} &= -\boldsymbol{x}_i^\top [1, 0, -1, 1, 0, -1, \cdots]^\top\\
    u(\boldsymbol{x}_i) &= \left| \frac{x_i^1 - \bar{x}^1}{\sigma^1} \right| + \cdots + \left| \frac{x_i^m - \bar{x}^m}{\sigma^m} \right| , \\
    \varepsilon_i &\sim \mathcal{N}(0,0.1),
\end{align*}
where $\bar{x}^j$ and $\sigma^j$ are the average and standard deviation of covariate $j$, respectively, in the dataset.
Note that $h(\boldsymbol{x}_i)$ and $\eta_i$ were not defined because the treatments were included in the dataset.

\begin{figure}[tbp]
        \begin{minipage}[t]{\columnwidth}
            \centering
            \includegraphics[width=0.8\linewidth]{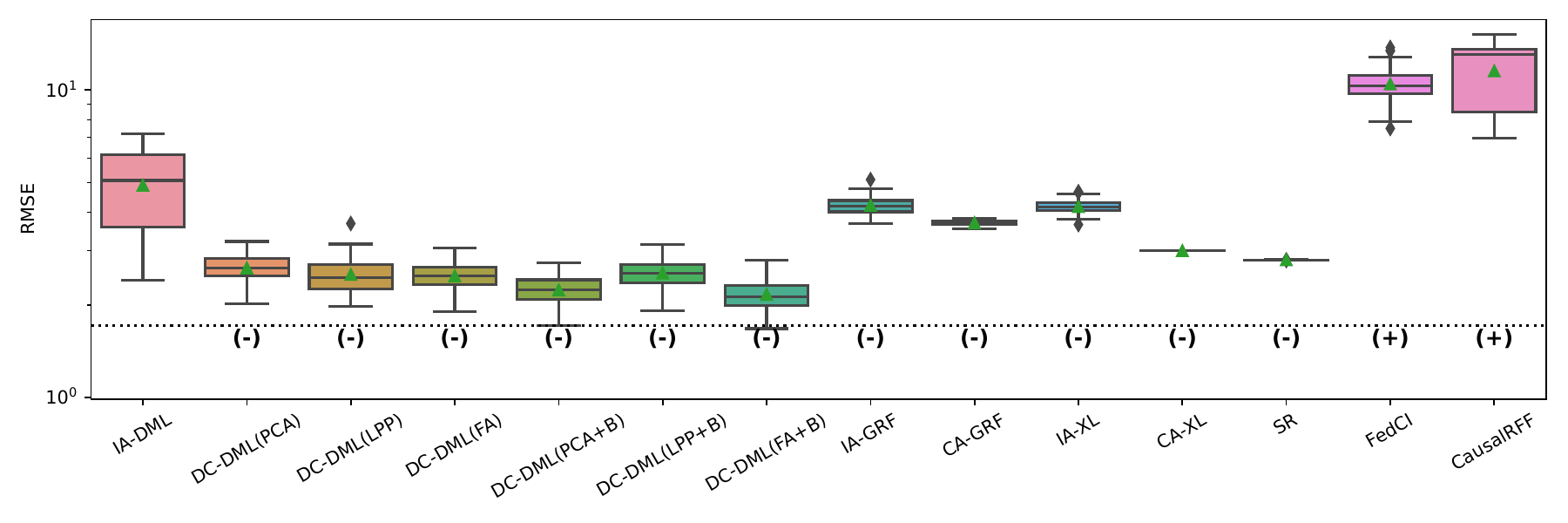}
            \subcaption{The RMSE of CATE}
            \label{fig:exp2_predrmse}
        \end{minipage}\\
        \begin{minipage}[t]{\columnwidth}
            \centering
            \includegraphics[width=0.75\linewidth]{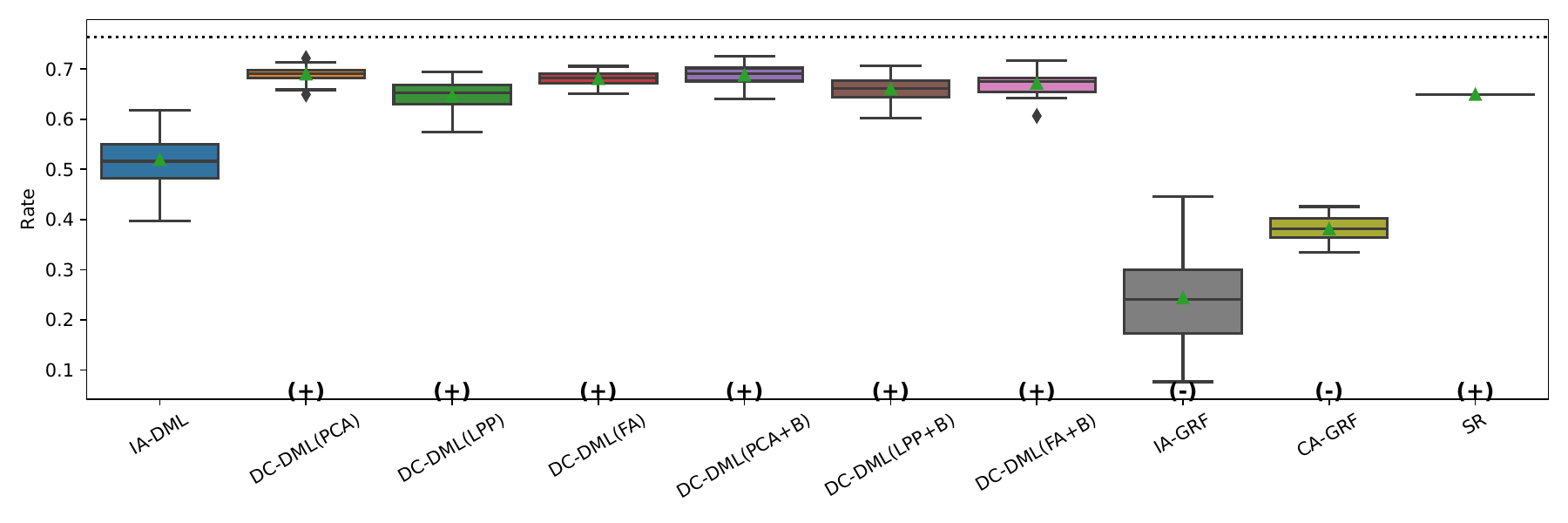}
            \subcaption{The consistency rate of significance of CATE}
            \label{fig:exp2_predsign}
        \end{minipage} \\
        \begin{minipage}[t]{\columnwidth}
            \centering
            \includegraphics[width=0.75\linewidth]{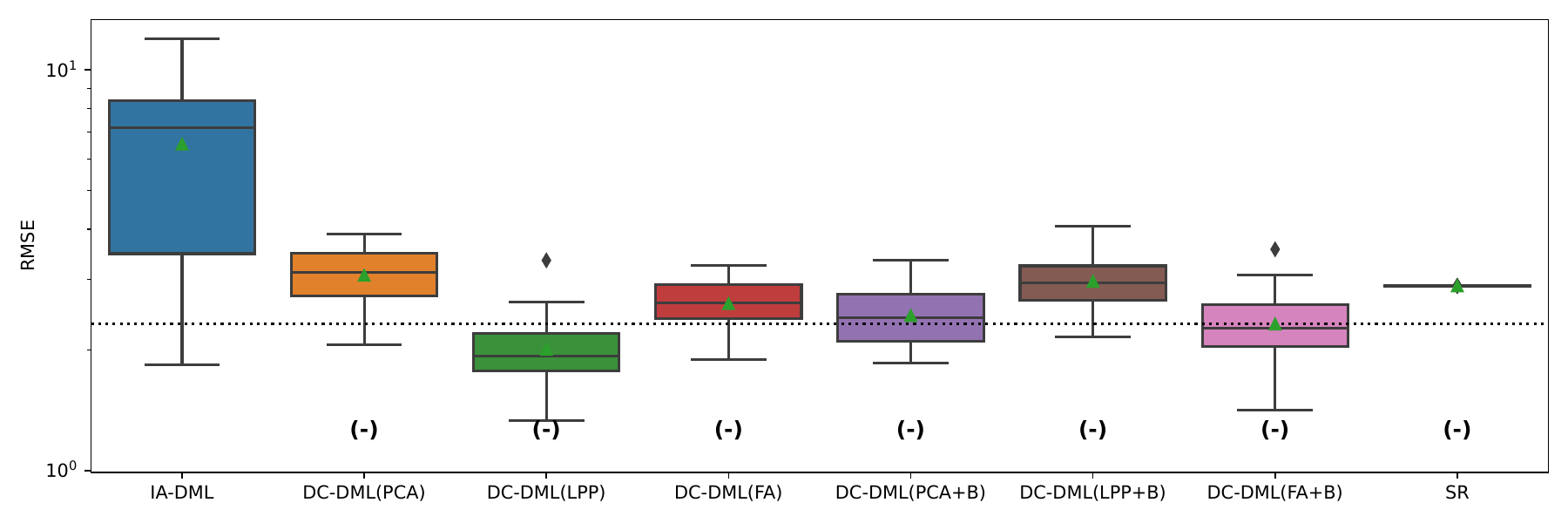}
            \subcaption{The RMSE of coefficients}
            \label{fig:exp2_coefrmse}
        \end{minipage} \\
        \begin{minipage}[t]{\columnwidth}
            \centering
            \includegraphics[width=0.75\linewidth]{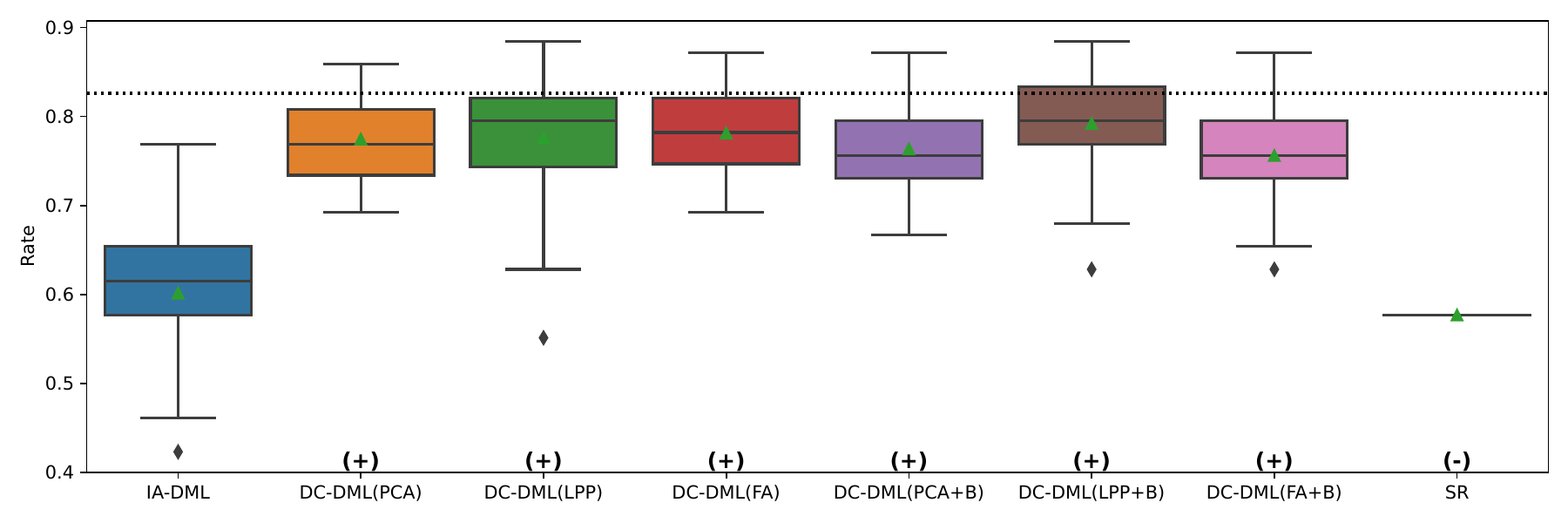}
            \subcaption{The consistency rate of significance of coefficients}
            \label{fig:exp2_coefsign}
        \end{minipage} \\     
    \caption{Results of Simulation II for CATE estimation. The dashed line is the average of CA-DML results. The triangle markers represent their average values. The vertical axes in Fig. \ref{fig:exp2_predrmse} and \ref{fig:exp2_coefrmse} are the logarithmic scales.}
    \label{fig:exp2}
\end{figure}

The results of the four evaluation measures in Simulation II are shown in Fig. \ref{fig:exp2_predrmse}-\ref{fig:exp2_coefsign} and Table \ref{tab:exp2_ate}.
In this simulation, we conducted 50 simulations with different random seed values for data distribution and machine learning, except for CausalRFF, which was conducted 10 times owing to its high computational cost.
The symbols $(+)$, $(-)$, and $(0)$ in the figures and table indicate the statistical significance of the differences between each method and IA-DML, based on a $t$-test: $(+)$ indicates a significantly positive difference, $(-)$ a significantly negative difference, and $(0)$ no significant difference.

\begin{table}[tb]
\centering
    \begin{tabular}{llllllllr}
    \hline \hline
    \multicolumn{1}{c}{} & \multicolumn{1}{c}{ave.} & \multicolumn{1}{c}{std.} & \multicolumn{1}{c}{min.} & \multicolumn{1}{c}{25\%} & \multicolumn{1}{c}{50\%} & \multicolumn{1}{c}{75\%} & \multicolumn{1}{c}{max.} & \multicolumn{1}{l}{\begin{tabular}[c]{@{}l@{}}significant\\ for IA-DML\end{tabular}} \\ \hline
    IA-DML & -1.3825 & 0.5513 & -2.5016 & -1.7629 & -1.3597 & -1.0049 & -0.0489 & \multicolumn{1}{l}{} \\
    DC-DML(PCA) & -0.9068 & 0.0868 & -1.1504 & -0.9605 & -0.9045 & -0.8685 & -0.5875 & (+) \\
    DC-DML(LPP) & -0.9163 & 0.1362 & -1.2067 & -0.9973 & -0.9358 & -0.8039 & -0.6636 & (+) \\
    DC-DML(FA) & -0.8361 & 0.1509 & -1.1914 & -0.9378 & -0.8665 & -0.7662 & -0.4295 & (+) \\
    DC-DML(PCA+B) & -0.7772 & 0.1248 & -1.1226 & -0.8504 & -0.7823 & -0.6897 & -0.5066 & (+) \\
    DC-DML(LPP+B) & -0.9371 & 0.1437 & -1.2871 & -1.0408 & -0.9246 & -0.8595 & -0.5926 & (+) \\
    DC-DML(FA+B) & -0.5204 & 0.2087 & -1.0854 & -0.6249 & -0.5008 & -0.3710 & 0.0646 & (+) \\
    IA-GRF & -1.2293 & 0.5937 & -2.7978 & -1.5681 & -1.2670 & -0.7801 & -0.1861 & (0) \\
    CA-GRF & -1.1965 & 0.1054 & -1.4081 & -1.2723 & -1.2023 & -1.1211 & -0.9440 & (+) \\
    IA-XL & -0.9099 & 0.5462 & -1.9572 & -1.2112 & -1.0067 & -0.5766 & 0.2795 & (+) \\
    CA-XL & -0.6120 & 0.0016 & -0.6163 & -0.6129 & -0.6122 & -0.6112 & -0.6078 & (+) \\
    SR & -0.4349 & 0.0000 & -0.4349 & -0.4349 & -0.4349 & -0.4349 & -0.4349 & (+) \\
    FedCI & -6.4676 & 1.8653 & -10.8365 & -7.2858 & -6.5230 & -5.3223 & -2.1092 & (-) \\
    CausalRFF & -7.0326 & 2.6672 & -10.4654 & -8.9366 & -7.3258 & -5.4126 & -2.3656 & (-) \\
    FIM & -0.7081 & 0.0701 & -0.8984 & -0.7383 & -0.6934 & -0.6666 & -0.5858 & (+) \\
    DC-QE & -0.6709 & 0.1385 & -1.0433 & -0.7525 & -0.6846 & -0.5980 & -0.3196 & (+) \\
    \hline \hline
    \multicolumn{9}{l}{True ATE $\approx 5.5883 \times 10^{-16}$} \\
    \multicolumn{9}{l}{Difference of average outcomes between the treatment and control groups $\approx -1.4704$} \\
    \multicolumn{9}{l}{Average of ATEs in CA-DML $\approx -0.5295$}
    \end{tabular}
\caption{The result of Simulation II for ATE estimation.}
\label{tab:exp2_ate}
\end{table}

As described below, Fig. \ref{fig:exp2_predrmse}-\ref{fig:exp2_coefsign} showed that DC-DML performed well for each of the evaluation measures in Simulation II.
First, as shown in Fig. \ref{fig:exp2_predrmse}, the result for the RMSE of CATE showed that DC-DML significantly outperformed IA-DML.
Moreover, DC-DML obtained better results than the existing methods.
Second, as shown in Fig. \ref{fig:exp2_predsign}, the result for the consistency rate of significance of CATE showed that DC-DML significantly outperformed IA-DML.
Moreover, DC-DML obtained better results than the existing methods.
Third, as shown in Fig. \ref{fig:exp2_coefrmse}, the result for the RMSE of coefficients showed that DC-DML significantly outperformed IA-DML and was comparable to CA-DML and SR.
Finally, as shown in Fig. \ref{fig:exp2_coefsign}, the result for the consistency rate of significance of coefficients showed that DC-DML significantly outperformed IA-DML.
Moreover, DC-DML obtained better results than SR.

Table \ref{tab:exp2_ate} showed that DC-DML outperformed IA-DML in terms of the ATE metric.
The true value of the ATE was approximately zero ($\approx 5.5883 \times 10^{-16}$), and DC-DML achieved estimates that were closer to this value than those obtained by IA-DML, indicating its higher estimation accuracy.
Although SR achieved the best performance, the second-best results were obtained by DC-DML (FA+B), suggesting that DC-DML was also competitive with other methods for ATE estimation.
In contrast, methods such as IA-GRF, FedCI, and CausalRFF did not perform as well as IA-DML.

These results from Simulation II indicated that DC-DML consistently outperformed IA-DML in CATE estimation, achieving lower RMSE and higher consistency in significance across all related metrics.
Compared to existing methods, DC-DML also showed competitive or better performance.
Although ATE estimation is an auxiliary focus, DC-DML still yielded reasonable results, further supporting its overall robustness.

\subsection{Simulation III: Robustness to distribution settings in real-world data}
\label{sec:simulation-robst}

Based on real-world datasets of financial assets and jobs, in Simulation III, we investigated the robustness of DC-DML performance in the case where the number of subjects or rate of treatment group across parties was imbalanced.
As the financial assets dataset, we focused on the survey of income and program participation (SIPP) dataset, which was used by Chernozhukov et al. \citep{chernozhukov2018double} to estimate the effect of 401(k) eligibility on accumulated assets.
\footnote{This is avaiable from the repository of \cite{chernozhukov2018double} (\url{https://github.com/VC2015/DMLonGitHub/}).}
The SIPP dataset consists of 9915 observations and included nine subject information (age, inc, educ, fsize, marr, twoearn, db, pira, hown) as covariates, 401(k) offerings as treatments, and net financial assets as outcomes.
As the jobs dataset, we focused on the Dehejia and Wahba dataset \citep{dehejia1999causal}, which was used to estimate the treatment effects of work experience and counseling on incomes.
\footnote{This is avaiable from the author's website (\url{https://users.nber.org/~rdehejia/nswdata.html}).}
The jobs dataset consists of 2675 observations and included eight subject information (age, black, hispanic, married, no degree, re74) as covariates, trainings as treatments, and real earnings in 1978 as outcomes.

In Simulation III, we considered three parties ($c = 3$) and three settings for data distribution where the rate of the treatment group and the number of subjects were even or uneven, as shown in Table \ref{tab:exp3_settings}.
In setting A, all parties had the same number of subjects and the same rate of treatment groups.
In setting B, the number of subjects was even, but the rate of treatment groups differed across parties.
In setting C, the number of subjects differed, but the rate of treatment groups was even across parties.

\begin{table}[tb]
\centering
    \scalebox{0.8}{
        \begin{tabular}{ll|rrr|rrr|rrr}
        \hline \hline
        \multicolumn{2}{l|}{} & \multicolumn{3}{l|}{setting A} & \multicolumn{3}{l|}{setting B} & \multicolumn{3}{l}{setting C} \\ \hline
        \multicolumn{2}{l|}{subjects} & \multicolumn{3}{l|}{even} & \multicolumn{3}{l|}{even} & \multicolumn{3}{l}{uneven} \\
        \multicolumn{2}{l|}{treatment rate} & \multicolumn{3}{l|}{even} & \multicolumn{3}{l|}{uneven} & \multicolumn{3}{l}{even} \\ \hline
        \multicolumn{1}{l|}{dataset} & party & \multicolumn{1}{l}{all} & \multicolumn{1}{l}{treated} & \multicolumn{1}{l|}{controlled} & \multicolumn{1}{l}{all} & \multicolumn{1}{l}{treated} & \multicolumn{1}{l|}{controlled} & \multicolumn{1}{l}{all} & \multicolumn{1}{l}{treated} & \multicolumn{1}{l}{controlled} \\ \hline
        \multicolumn{1}{l|}{\multirow{3}{*}{financial assets}} & 1 & 3304 & 1227 & 2077 & 3304 & 2549 & 755 & 6864 & 2549 & 4315 \\
        \multicolumn{1}{l|}{} & 2 & 3304 & 1227 & 2077 & 3304 & 849 & 2455 & 2287 & 849 & 1438 \\
        \multicolumn{1}{l|}{} & 3 & 3304 & 1227 & 2077 & 3304 & 283 & 3021 & 762 & 283 & 479 \\ \hline
        \multicolumn{1}{l|}{\multirow{3}{*}{jobs}} & 1 & 891 & 61 & 830 & 891 & 92 & 799 & 1337 & 92 & 1245 \\
        \multicolumn{1}{l|}{} & 2 & 891 & 61 & 830 & 891 & 61 & 830 & 891 & 61 & 830 \\
        \multicolumn{1}{l|}{} & 3 & 891 & 61 & 830 & 891 & 30 & 861 & 445 & 30 & 415 \\ \hline \hline
        \end{tabular}
    }
\caption{The number of subjects in each setting for Simulation III.}
\label{tab:exp3_settings}
\end{table}

In Simulation III, the result of CA-DML was used as the benchmark.
Table  \ref{tab:exp3_benchmarks} showed the average results of 50 CA-DML trials.
Note that comparisons with GRF, X-Learner, and FedCI did not make sense because those assume the non-linear CATE model.
Thus, only CA-DML, IA-DML and SR were considered for comparison with DC-DML.

\begin{table}[tb]
\centering
    \begin{tabular}{lrllrl}
    \hline \hline
    \multicolumn{3}{l|}{financial   assets} & \multicolumn{3}{l}{jobs} \\ \hline
    covariate & \multicolumn{2}{l|}{coefficient} & covariate & \multicolumn{2}{l}{coefficient} \\ \hline
    const. & -9705.1794 & \multicolumn{1}{l|}{} & const. & -7533.7519 &  \\
    age & 172.0307 & \multicolumn{1}{l|}{} & age & 226.4912 & * \\
    inc & -0.1297 & \multicolumn{1}{l|}{} & black & 2493.7939 &  \\
    educ & 642.9040 & \multicolumn{1}{l|}{} & hispanic & 1519.3141 &  \\
    fsize & -1003.0686 & \multicolumn{1}{l|}{} & married & -2215.7011 &  \\
    marr & 1102.9090 & \multicolumn{1}{l|}{} & no\_degree & -858.5624 &  \\
    twoearn & 5607.3989 & \multicolumn{1}{l|}{} & re74 & -0.3354 &  \\
    db & 5657.7264 & \multicolumn{1}{l|}{*} &  &  &  \\
    pira & -1032.3599 & \multicolumn{1}{l|}{} &  &  &  \\
    hown & 5324.2854 & \multicolumn{1}{l|}{*} &  &  &  \\ \hline \hline
    \multicolumn{3}{l}{* $p<0.05$} &  & \multicolumn{1}{l}{} & 
    \end{tabular}
\caption{The benchmark coefficients of Simulation III.}
\label{tab:exp3_benchmarks}
\end{table}

\begin{figure}[tb]
  \centering
  \includegraphics[width=0.75\linewidth]{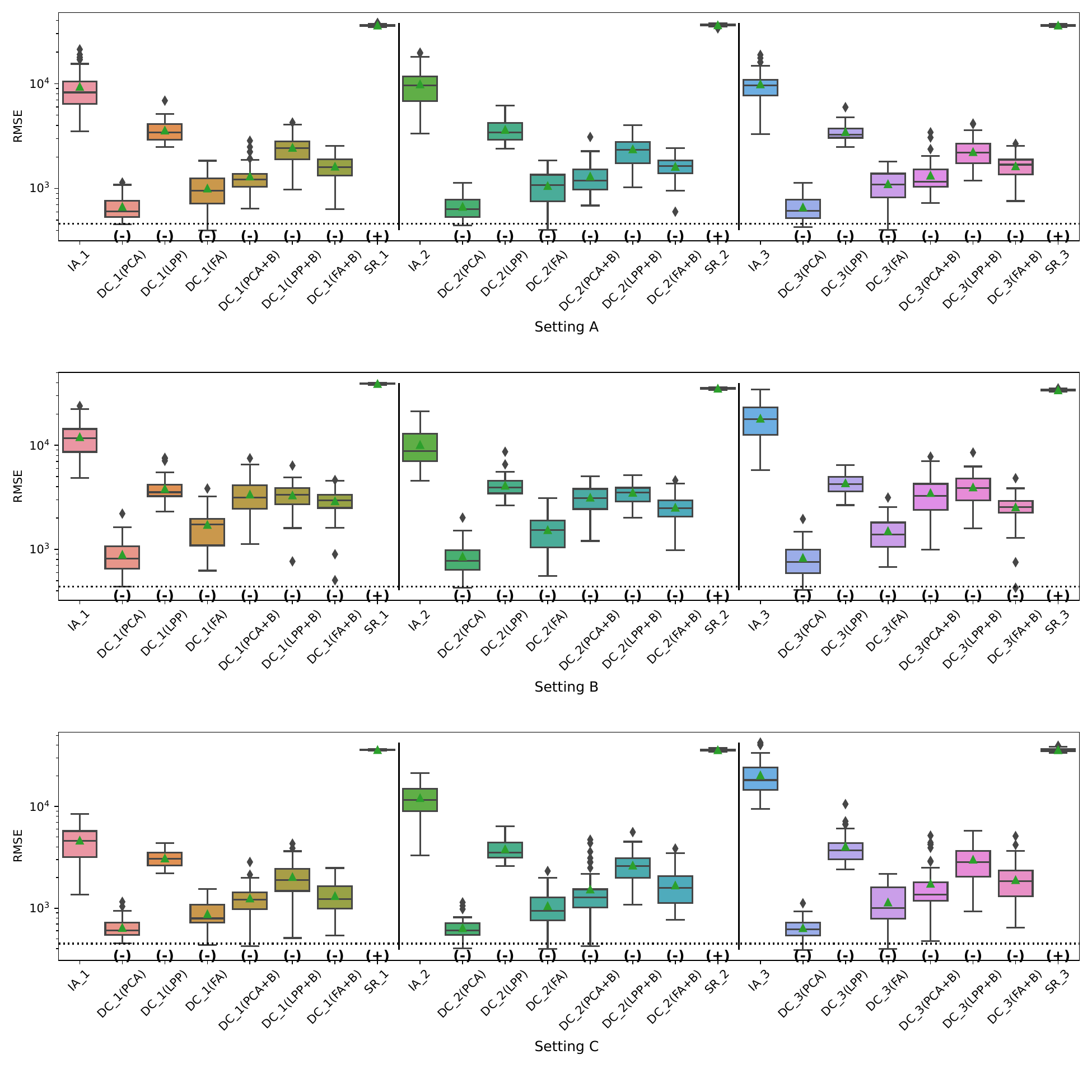}
  \caption{The RMSE of CATE in Simulation III for the financial assets dataset. The dashed line is the average of CA-DML results. The triangle markers represent their average values. The vertical axis is the logarithmic scale.}
  \label{fig:exp3_sipp_predrmse}
\end{figure}

Fig. \ref{fig:exp3_sipp_predrmse} showed the RMSE of CATE in Simulation III for the financial assets dataset.
The symbols $(+)$, $(-)$ and $(0)$ in the figures represent significant positive, significant negative and not significant differences from the IA-DML results, respectively, in $t$-test.
The results for parties 1, 2 and 3 are shown on the left, middle and right sides, respectively, in the figures.
In all the cases shown in Fig. \ref{fig:exp3_sipp_predrmse}, DC-DML results were better than IA-DML and SR.
In addition, the dimensionality reduction method combined with bootstrapping showed robust results.
However, DC-DML using LPP was relatively worse than for DC-DML using the other dimensionality reduction methods.
No SR results were better than IA-DML.

\begin{figure}[tb]
  \centering
  \includegraphics[width=0.75\linewidth]{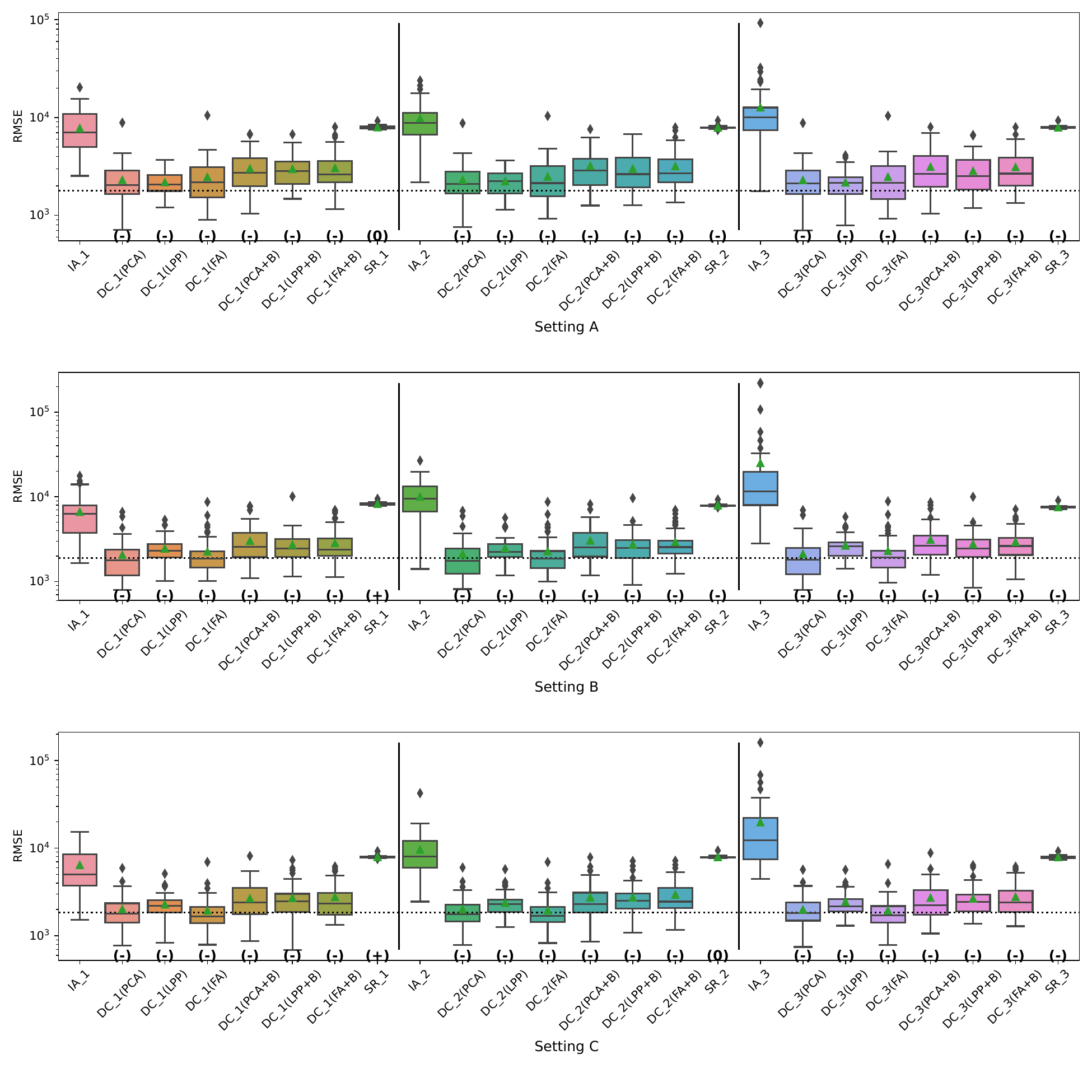}
  \caption{The RMSE of CATE in Simulation III for the jobs dataset. The meaning of symbols, lines and axes are the same with Fig. \ref{fig:exp3_sipp_predrmse}.}
  \label{fig:exp3_jobs_predrmse}
\end{figure}

Fig. \ref{fig:exp3_jobs_predrmse} showed the RMSE of CATE in Simulation III for the jobs dataset.
Similarly to the simulation for the financial assets, DC-DML results were better than IA-DML and SR in all cases in Fig. \ref{fig:exp3_jobs_predrmse}.
Although, as shown in \ref{app:expIII}, DC-DML did not always exhibit better performance than IA-DML in the other measures, the results of DC-DML were often better than those of IA-DML.
As with the financial assets dataset, the dimensionality reduction method combined with bootstrapping showed robust results, and SR results were often worse than IA-DML.

The simulation results in both datasets suggested that the choice of dimensionality reduction method had a significant impact on the performance of DC-DML.
What dimensionality reduction method provides good performance for DC-DML is likely to be data-dependent.
The question of how to select a dimensionality reduction method is a future issue.

In summary, in Simulation III with two datasets, DC-DML achieved better results than IA-DML, although not in all cases.
Those results suggest that DC-DML can obtain results comparable to CA-DML under various data distribution settings, unlike IA-DML.

\section{Discussion}
\label{sec:discussion}
Utilizing distributed data across multiple parties is an important topic in data science, but confidential or privacy concerns need to be addressed.
Multi-source data fusion to construct models for the estimation of CATE can help to provide better policy or medical treatments for subjects.
Our method provides statistical test results for treatment effects.
This allows us to make decisions such as applying the treatment only to subjects with significantly better effects, or discontinuing the treatment for subjects with significantly worse effects.
Giving procedures that take into account the subjects' covariates is essential for safer and more efficient policy and health care implementation.

Our method enables the collaborative representation to be constructed again for each additional collaboration party.
In addition, our method requires communication with the analyst only at the time of sharing intermediate representations or receiving estimation results, and does not require long time synchronization.
This advantage enables intermediate representations to be accumulated over a longer time involving a larger number of parties, and then CATE can be estimated in refined models with large accumulated data, i.e. knowledge base.

Here, we briefly discuss the relationship between the readily identifiability, an important issue in the privacy preservation literature, and our method.
In the DC analysis literature, the readily identifiability indicates that the original data containing privacy information can be readily identified with the intermediate representation \citep{imakura2023non}.
Solving the readily identifiability issue could enhance the privacy preservation of our method.
As shown in \ref{app:non-ri}, our method can be extended to use intermediate representations that are not readily identifiable with the original data.
For a more detailed discussion for the readily identifiability, see \ref{app:non-ri}.

Another theoretical aspect worth discussing is whether the convergence property of DML, particularly its robustness to slow convergence rates of nuisance parameter estimators under the Neyman--orthogonality condition, still holds in the distributed data setting.
If all assumptions of Theorem \ref{thm:correctness} are satisfied, the estimation procedure in DC-DML reduces to standard DML, and thus similar convergence guarantees are expected to hold.
However, in our setting, additional factors, such as the dimensionality reduction process, may influence the approximation quality of the collaborative representation.
Understanding how these factors affect the convergence property and theoretical guarantees is an important direction for future work.

We also note that in Simulation II, FedCI and CausalRFF did not perform as well as DC-DML. 
However, this result should be interpreted with caution.
First, the simulation setting assumes a linear CATE model, which aligns with the modeling assumption of DC-DML and thus may favor our method over more flexible approaches.
Second, although we used the default hyperparameters for FedCI and CausalRFF, we did not perform extensive hyperparameter tuning for these methods.
It is therefore possible that their performance could improve under more carefully optimized configurations.
A more comprehensive comparison across various causal structures and tuning strategies is an important direction for future work.

Our study has some limitations.
First, our method assumes the linear CATE model.
As mentioned in Section \ref{sec:method_ad_disad}, the linear CATE model may obtain incorrect conclusions in data where the assumption is not valid.
Second, it is not clear how the dimensionality reduction method should be selected in DC-DML.
Third, DC-DML relies on the accurate specification and estimation of the outcome and treatment prediction models, which may affect the validity of the estimated CATEs.
Addressing these issues is an important direction for future work.

\section{Conclusion}
\label{sec:conclusion}
In this study, we proposed DC-DML, which estimates CATEs from privacy-preserving fusion data constructed from distributed data.
Our method enables flexible estimation of CATE models from fusion data through a semi-parametric approach and does not require iterative communication.
Moreover, our method enables collaborative estimation between multiple time points and different parties through the accumulation of a knowledge base.
These are important characteristics with application to the practical use of our methods.
Through the three simulations, we demonstrated the effectiveness of our method as follows.
First, even in situations where it is difficult to make valid estimates using IA-DML, DC-DML can produce more valid estimates than IA-DML.
Second, DC-DML can outperform IA-DML and obtain equal or better results than the existing methods.
Third, DC-DML can obtain results comparable to those of CA-DML across various data distribution settings, outperforming IA-DML.

In the future, we will develop our model to address non-linear CATE models.
Other directions include extending our method to vertically distributed data, and conducting experiments on marketing datasets.


\section*{Acknowledgement}
The authors gratefully acknowledge the New Energy and Industrial Technology Development Organization (NEDO) (No. JPNP18010), Japan Science and Technology Agency (JST) (No. JPMJPF2017), Japan Society for the Promotion of Science (JSPS), and Grants-in-Aid for Scientific Research (Nos. JP19KK0255, JP21H03451, JP23K22166, JP22K19767, JP23K28192).

\section*{Declaration of generative AI and AI-assisted technologies in the writing process}
During the preparation of this work the authors used ChatGPT (OpenAI) to assist with phrasing, grammar correction, and refinement of English expressions in the manuscript. After using this tool, the authors reviewed and edited the content as needed and take full responsibility for the content of the publication.

\appendix

\section{Acronym table}
\label{app:acronym}
A list of acronyms and definitions used in this paper is shown in Table \ref{tab:acronym}.
\begin{table}[tb]
\centering
    \begin{tabular}{ll}
    \hline
    Acronym & Definition \\ \hline
    AIPW & Augmented inverse propensity weighted \\
    ATE & Average treatment effect \\
    CA & Centralized analysis \\
    CATE & Conditional average treatment effect \\
    CausalRFF & Causal random Fourier features \\
    DC & Data collaboration \\
    DC-DML & Data collaboration double machine learning \\
    DC-QE & Data collaboration quasi-experiment \\
    DML & Double machine learning \\
    FA & Factor analysis \\
    FedCI & Federated causal inference \\
    FIM & Federated IPW-ML \\
    GRF & Generalized random forest \\
    IA & Individual analysis \\
    IHDP & Infant health and development program \\
    IPW & Inverse propensity-weighted \\
    LGBM & Light gradient boosting machine \\
    LPP & Locality preserving projection \\
    MLE & Maximum likelihood estimator \\
    PCA & Principal component analysis \\
    RMSE & Root mean squared error \\
    SIPP & Survey of income and program participation \\
    SR & Secure regression \\
    SVM & Support vector machine \\ \hline
    \end{tabular}
\caption{Acronym table.}
\label{tab:acronym}
\end{table}

\section{Algorithm of data collaboration quasi-experiment}
\label{app:alg_deqe}
Algorithm \ref{alg:dcqe} is the pseudo-code of DC-QE.

\begin{algorithm}[tb]
    \caption{Data collaboration quasi-experiment (DC-QE)}
    \label{alg:dcqe}
    \begin{algorithmic}[1]
        \Statex \textbf{Input: covariates $X$, treatments $Z$, outcomes $Y$}.
        \Statex \textbf{Output: estimated treatment effect $\hat{\tau}$}.
        \vspace{-.3\baselineskip}
        
        \Statex \hrulefill
        \vspace{-.3\baselineskip}
        \Statex \textit{user-$k$-side}
        \vspace{-.5\baselineskip}
        \Statex \hrulefill
        \vspace{-.1\baselineskip}
        \State Generate anchor dataset $X_{k}^\text{anc}$ and share it with all users.
        \State Set $X^\text{anc}$.
        \State Generate $f_{k}$.
        \State Compute $\widetilde{X}_{k} = f_{k}(X_{k})$.
        \State Compute $\widetilde{X}_{k}^\text{anc} = f_{k}(X^\text{anc})$.
        \State Share $\widetilde{X}_{k}$, $\widetilde{X}_{k}^\text{anc}$, $Z_k$ and $Y_k$ to the analyst.
        \vspace{-.3\baselineskip}
        
        \Statex \hrulefill
        \vspace{-.3\baselineskip}
        \Statex \textit{analyst-side}
        \vspace{-.5\baselineskip}
        \Statex \hrulefill
        \vspace{-.1\baselineskip}
        \State Get $\widetilde{X}_{k}$, $\widetilde{X}_{k}^\text{anc}$, $Z_k$ and $Y_k$ for all $k$.
        \State Set $\widetilde{X}_{k}$ and $\widetilde{X}_{k}^\text{anc}$.
        \State Compute $g_k$ from $\widetilde{X}_{k}^\text{anc}$ for all $k$ such that $g_k(\widetilde{X}_{k}^\text{anc}) \approx g_{k'}(\widetilde{X}_{k'}^\text{anc}) ~ (k \neq k')$.
        \State Compute $\check{X}_k = g_k(\widetilde{X}_{k})$ for all $k$.
        \State Set $\check{X}$, $Z$ and $Y$.
        \State Estimate propensity scores $\hat{\boldsymbol{\alpha}}$ from $\check{X}$ and $Z$.
        \State Estimate average treatment effect $\hat{\tau}$ from $\hat{\boldsymbol{\alpha}}$ and $Y$ using existing ways (matching, weighting and so on).

        \end{algorithmic}
\end{algorithm}

\section{The proof of Theorem \ref{thm:score}}
\label{app:proof_score}
\begin{proof}
    Denote $\bar{\boldsymbol{x}}_i = [1, \boldsymbol{x}_i^\top]^\top$.
    First, we can prove the moment condition as
    \begin{align*}
        \mathbb{E} [\psi(\boldsymbol{x}_i; \boldsymbol{\beta}, q, h)] &= 
        \mathbb{E} \left[ \bar{\boldsymbol{x}}_i (z_i - h(\boldsymbol{x}_i))(y_i - q(\boldsymbol{x}_i) - \bar{\boldsymbol{x}}_i^\top \boldsymbol{\beta} (z_i - h(\boldsymbol{x}_i))) \right] \\
        &= \mathbb{E} \left[ \bar{\boldsymbol{x}}_i \eta_i \varepsilon_i \right] \\
        &= \mathbb{E} \left[ \bar{\boldsymbol{x}}_i \mathbb{E} \left[ \eta_i \varepsilon_i | \boldsymbol{x}_i \right] \right] \\
        &= \boldsymbol{0}.
    \end{align*}
    Second, for the Neyman--orthogonality, consider
    \begin{align*}
        &\psi(\boldsymbol{x}_i; \boldsymbol{\beta}, q + r\delta_q, h + r\delta_h) \\
        &= \bar{\boldsymbol{x}}_i (z_i - h(\boldsymbol{x}_i) - r\delta_h(\boldsymbol{x}_i)) (y_i - q(\boldsymbol{x}_i) - r\delta_q(\boldsymbol{x}_i) - \bar{\boldsymbol{x}}_i^\top \boldsymbol{\beta} (z_i - h(\boldsymbol{x}_i) - r\delta_h(\boldsymbol{x}_i))) \\
        &= \bar{\boldsymbol{x}}_i (\eta_i - r\delta_h(\boldsymbol{x}_i)) (\zeta_i - r\delta_q(\boldsymbol{x}_i) - \bar{\boldsymbol{x}}_i^\top \boldsymbol{\beta} (\eta_i - r\delta_h(\boldsymbol{x}_i))) \\
        &= \bar{\boldsymbol{x}}_i \left\{ (\eta_i - r\delta_h(\boldsymbol{x}_i)) (\eta_i - r\delta_q(\boldsymbol{x}_i)) - \bar{\boldsymbol{x}}_i^\top \boldsymbol{\beta} (\eta_i - r\delta_h(\boldsymbol{x}_i))^2 \right\},
    \end{align*}
    where $\delta_h = h'-h$ and $\delta_q = q'-q$.
    Then,
    \begin{align*}
        &\lim_{r \to 0} \frac{\partial}{\partial r} \mathbb{E} \left[ \psi(\boldsymbol{x}_i; \boldsymbol{\beta}, q + r\delta_q, h + r\delta_h) \right] \\
        &= \lim_{r \to 0} \frac{\partial}{\partial r} \mathbb{E} \left[ \bar{\boldsymbol{x}}_i \mathbb{E} \left[ \left\{ (\eta_i - r\delta_h(\boldsymbol{x}_i)) (\zeta_i - r\delta_q(\boldsymbol{x}_i)) - \bar{\boldsymbol{x}}_i^\top \boldsymbol{\beta} (\eta_i - r\delta_h(\boldsymbol{x}_i))^2 \right\} | \boldsymbol{x}_i \right] \right] \\
        &= \mathbb{E} \left[ \bar{\boldsymbol{x}}_i \lim_{r \to 0} \frac{\partial}{\partial r} \mathbb{E} \left[ \left\{ (\eta_i - r\delta_h(\boldsymbol{x}_i)) (\zeta_i - r\delta_q(\boldsymbol{x}_i)) - \bar{\boldsymbol{x}}_i^\top \boldsymbol{\beta} (\eta_i - r\delta_h(\boldsymbol{x}_i))^2 \right\} | \boldsymbol{x}_i \right] \right].
    \end{align*}
    Here,
    \begin{align*}
        &\lim_{r \to 0} \frac{\partial}{\partial r} \mathbb{E} \left[ \left\{ (\eta_i - r\delta_h(\boldsymbol{x}_i)) (\zeta_i - r\delta_q(\boldsymbol{x}_i)) - \bar{\boldsymbol{x}}_i^\top \boldsymbol{\beta} (\eta_i - r\delta_h(\boldsymbol{x}_i))^2 \right\} | \boldsymbol{x}_i \right] \\
        &= \lim_{r \to 0} \frac{\partial}{\partial r} \mathbb{E} \left[ (\eta_i - r\delta_h(\boldsymbol{x}_i)) (\zeta_i - r\delta_q(\boldsymbol{x}_i)) | \boldsymbol{x}_i \right]
        - \lim_{r \to 0} \frac{\partial}{\partial r} \mathbb{E} \left[ \bar{\boldsymbol{x}}_i^\top \boldsymbol{\beta} (\eta_i - r\delta_h(\boldsymbol{x}_i))^2 | \boldsymbol{x}_i \right].
    \end{align*}
    The first term is
    \begin{align*}
        &\lim_{r \to 0} \frac{\partial}{\partial r} \mathbb{E} \left[ (\eta_i - r\delta_h(\boldsymbol{x}_i)) (\zeta_i - r\delta_q(\boldsymbol{x}_i)) | \boldsymbol{x}_i \right]\\
        &= \lim_{r \to 0} \mathbb{E} \left[  - \eta_i \delta_q(\boldsymbol{x}_i) - \zeta_i \delta_h(\boldsymbol{x}_i) + 2r \delta_h(\boldsymbol{x}_i) \delta_q(\boldsymbol{x}_i) | \boldsymbol{x}_i \right] \\
        &= \mathbb{E} \left[ - \eta_i \delta_q(\boldsymbol{x}_i) - \zeta_i \delta_h(\boldsymbol{x}_i) | \boldsymbol{x}_i \right] \\
        &= - \delta_q(\boldsymbol{x}_i) \mathbb{E} \left[ \eta_i | \boldsymbol{x}_i \right] - \delta_h(\boldsymbol{x}_i) \mathbb{E} \left[ \zeta_i | \boldsymbol{x}_i \right] \\
        &= \boldsymbol{0},
    \end{align*}
    and the second term is
    \begin{align*}
        &\lim_{r \to 0} \frac{\partial}{\partial r} \mathbb{E} \left[ \bar{\boldsymbol{x}}_i^\top \boldsymbol{\beta} (\eta_i - r\delta_h(\boldsymbol{x}_i))^2 | \boldsymbol{x}_i \right] \\
        &= \bar{\boldsymbol{x}}_i^\top \boldsymbol{\beta} \lim_{r \to 0} \frac{\partial}{\partial r} \mathbb{E} \left[ (\eta_i - r\delta_h(\boldsymbol{x}_i))^2 | \boldsymbol{x}_i \right] \\
        &= \bar{\boldsymbol{x}}_i^\top \boldsymbol{\beta} \lim_{r \to 0} \mathbb{E} \left[ -2 (\eta_i - r\delta_h(\boldsymbol{x}_i))\delta_h(\boldsymbol{x}_i) | \boldsymbol{x}_i \right] \\
        &= -2 \bar{\boldsymbol{x}}_i^\top \boldsymbol{\beta} \delta_h(\boldsymbol{x}_i) \lim_{r \to 0} \mathbb{E} \left[(\eta_i - r\delta_h(\boldsymbol{x}_i)) | \boldsymbol{x}_i \right] \\
        &= -2 \bar{\boldsymbol{x}}_i^\top \boldsymbol{\beta} \delta_h(\boldsymbol{x}_i) \mathbb{E} \left[\eta_i | \boldsymbol{x}_i \right] \\
        &= \boldsymbol{0}.
    \end{align*}
    Thus,
    \begin{equation*}
        \lim_{r \to 0} \frac{\partial}{\partial r} \mathbb{E} \left[ \psi(\boldsymbol{x}_i; \boldsymbol{\beta}, q + r\delta_q, h + r\delta_h) \right] = \boldsymbol{0}.
    \end{equation*}
\end{proof}

\section{The proof of Theorem \ref{thm:correctness}}
\label{app:proof_correctness}
\begin{proof}
    From Theorem 1 in \cite{imakura2021accuracy}, if
    \begin{equation*}
        \mathcal{S}_1 = \cdots = \mathcal{S}_c ~ \textup{and} ~ \boldsymbol{\mu}_1 = \cdots = \boldsymbol{\mu}_c ~ \textup{and} ~ \textup{rank}(X^{\textup{anc}} \bar{F}_k) = \widetilde{m},
    \end{equation*}
    then we have
    \begin{equation*}
        \check{X} = W \bar{F_1} G_1 = \cdots = W \bar{F_c} G_c,
    \end{equation*}
    where $W = [\boldsymbol{1} , X - \boldsymbol{1} \boldsymbol{\mu}_1^\top]$.
    Denote $\boldsymbol{w}_i^\top$ as the $i$th row of $W$.
    Here, consider
    \begin{equation*}
        t(\boldsymbol{\beta}^{\prime}) = \left[ [1, \mu_1^1, \cdots, \mu_1^m] \boldsymbol{\beta}^{\prime}, \beta^{\prime 1}, \cdots, \beta^{\prime m} \right]^\top \in \mathbb{R}^{m+1}.
    \end{equation*}
    then, we can express $\boldsymbol{\gamma}_{\textup{CA}} = t(\boldsymbol{\beta}_{\textup{CA}})$.
    Since $[1, \boldsymbol{x}_i^\top] \boldsymbol{\beta}^{\prime} = \boldsymbol{w}_i^\top t(\boldsymbol{\beta}^{\prime})$,
    \begin{align*}
        \boldsymbol{\beta}_{\textup{CA}} &= \argmin_{\boldsymbol{\beta}^{\prime} \in \mathbb{R}^{m+1}} \sum_{i} (\hat{\zeta}_i - \hat{\eta}_i [1, \boldsymbol{x}_i^\top] \boldsymbol{\beta}^{\prime})^2 \\
        &= \argmin_{\boldsymbol{\beta}^{\prime} \in \mathbb{R}^{m+1}} \sum_{i} (\hat{\zeta}_i - \hat{\eta}_i \boldsymbol{w}_i^\top t(\boldsymbol{\beta}^{\prime}))^2.
    \end{align*}
    Thus, $\boldsymbol{\gamma}_{\textup{CA}}$ is equal to the solution of the following least squares problem:
    \begin{equation*}
        \boldsymbol{\gamma}_{\textup{CA}} = \argmin_{\boldsymbol{\gamma}^{\prime} \in \mathbb{R}^{m+1}} \sum_{i} (\hat{\zeta}_i - \hat{\eta}_i \boldsymbol{w}_i^\top \boldsymbol{\gamma}^{\prime})^2.
    \end{equation*}
    From the assumption $\boldsymbol{\gamma}_{\textup{CA}} \in \mathcal{S}_k$,
    \begin{align*}
        \boldsymbol{\gamma_k}   &= \bar{F_k} G_k \check{\boldsymbol{\gamma}} \\
                                &= \argmin_{\bar{F_k} G_k \boldsymbol{\gamma}^{\prime} \in \mathcal{S}_k} \sum_{i} (\hat{\zeta}_i - \hat{\eta}_i \boldsymbol{w}_i^\top \bar{F_k} G_k \boldsymbol{\gamma}^{\prime})^2 \\
                                &= \argmin_{\boldsymbol{\gamma}^{\prime} \in \mathcal{S}_k} \sum_{i} (\hat{\zeta}_i - \hat{\eta}_i \boldsymbol{w}_i^\top \boldsymbol{\gamma}^{\prime})^2 \\
                                &= \boldsymbol{\gamma}_{\textup{CA}}.
    \end{align*}
    We used the assumptions $q_{\text{CA}}(\boldsymbol{x}_i) = \check{q}(\check{\boldsymbol{x}}_i)$ and $h_{\text{CA}}(\boldsymbol{x}_i) = \check{h}(\check{\boldsymbol{x}}_i)$ to transform the equation from line 2 to 3.
    It is obvious that $\boldsymbol{\beta}_{\text{CA}}$ and $\boldsymbol{\beta}_k$ are equal in all but the first element.
    However,
    \begin{align*}
        \alpha_k    &= [1, -\mu_k^1, \cdots, -\mu_k^m] \boldsymbol{\gamma}_k \\
                    &= [1, -\mu_1^1, \cdots, -\mu_1^m] \boldsymbol{\gamma}_{\text{CA}} \\
                    &= \alpha_{\text{CA}}
    \end{align*}
    where $\alpha_{\text{CA}}$ is the first element of $\boldsymbol{\beta}_{\text{CA}}$.
    Then, $\boldsymbol{\beta}_{\text{CA}} = \boldsymbol{\beta}_k$.
\end{proof}

\section{Hyperparameters for candidate methods}
\label{app:hparams_cand}
Most hyperparameters of candidate methods for $q$ and $h$ are default parameters in scikit-learn (V1.2.2) and LightGBM (V3.3.5).
The hyperparameters for $q$ estimation are as follows.
\begin{description}
    \item[LinearRegression (scikit-learn)] fit\_intercept: True, positive: False
    \item[RandomForestRegressor (scikit-learn)] bootstrap: True, ccp\_alpha: 0.0, criterion: squared\_error, max\_depth: None, max\_features: 1.0, max\_leaf\_nodes: None, max\_samples: None, min\_impurity\_decrease: 0.0, min\_samples\_leaf: 1, min\_samples\_split: 2, min\_weight\_fraction\_leaf: 0.0, n\_estimators: 100, oob\_score: False, warm\_start: False
    \item[KNeighborsRegressor (scikit-learn)] algorithm: auto, leaf\_size: 30, metric: minkowski, metric\_params: None, n\_neighbors: 5, p: 2, weights: uniform
    \item[LGBMRegressor (LightGBM)] boosting\_type: gbdt, class\_weight: None, colsample\_bytree: 1.0, importance\_type: split, learning\_rate: 0.1, max\_depth: -1, min\_child\_samples: 20, min\_child\_weight: 0.001, min\_split\_gain: 0.0, n\_estimators: 100, num\_leaves: 31, objective: None, reg\_alpha: 0.0, reg\_lambda: 0.0, subsample: 1.0, subsample\_for\_bin: 200000, subsample\_freq: 0
    \item[SVR (scikit-learn)] C: 1.0, cache\_size: 200, coef0: 0.0, degree: 3, epsilon: 0.1, gamma: scale, kernel: rbf, max\_iter: -1, shrinking: True, tol: 0.001
\end{description}
The hyperparameters for $h$ estimation are as follows.
\begin{description}
    \item[LogisticRegression (scikit-learn)] C: 1.0, class\_weight: None, dual: False, fit\_intercept: False, intercept\_scaling: 1, l1\_ratio: None, max\_iter: 100, multi\_class: auto, n\_jobs: None, penalty: none, solver: lbfgs, tol: 0.0001, warm\_start: False
    \item[RandomForestClassifier (scikit-learn)] bootstrap: True, ccp\_alpha: 0.0, class\_weight: None, criterion: gini, max\_depth: None, max\_features: sqrt, max\_leaf\_nodes: None, max\_samples: None, min\_impurity\_decrease: 0.0, min\_samples\_leaf: 1, min\_samples\_split: 2, min\_weight\_fraction\_leaf: 0.0, n\_estimators: 100, oob\_score: False, warm\_start: False
    \item[KNeighborsClassifier (scikit-learn)] algorithm: auto, leaf\_size: 30, metric: minkowski, metric\_params: None, n\_neighbors: 5, p: 2, weights: uniform
    \item[LGBMClassifier (LightGBM)] boosting\_type: gbdt, class\_weight: None, colsample\_bytree: 1.0, importance\_type: split, learning\_rate: 0.1, max\_depth: -1, min\_child\_samples: 20, min\_child\_weight: 0.001, min\_split\_gain: 0.0, n\_estimators: 100, num\_leaves: 31, objective: None, reg\_alpha: 0.0, reg\_lambda: 0.0, subsample: 1.0, subsample\_for\_bin: 200000, subsample\_freq: 0
    \item[SVC (scikit-learn)] C: 1.0, break\_ties: False, cache\_size: 200, class\_weight: None, coef0: 0.0, decision\_function\_shape: ovr, degree: 3, gamma: scale, kernel: rbf, max\_iter: -1, probability: True, shrinking: True, tol: 0.001
\end{description}

\section{The results of the preliminary simulations}
\label{app:pre_exp}
In the preliminary simulations, RMSE was calculated for the estimation of $q$ and the Brier score for the estimation of $h$ using a two-fold cross-validation with raw data.
Table \ref{tab:preexp} showed the averages of 50 trials.

\begin{table}[tb]
\centering
    \begin{tabular}{lrlr}
    \hline \hline
    \multicolumn{4}{c}{Simulation II} \\ \hline
    $q$ & \multicolumn{1}{l|}{RMSE} & $h$ & \multicolumn{1}{l}{Brier score} \\ \hline
    Multiple regression & \multicolumn{1}{r|}{2.5613} & Logistic regression & 0.2579 \\
    Random forests & \multicolumn{1}{r|}{2.5308} & Random forests & 0.1532 \\
    K-nearest neighbor & \multicolumn{1}{r|}{2.8690} & K-nearest neighbor & 0.1665 \\
    LGBM & \multicolumn{1}{r|}{2.5734} & LGBM & 0.1733 \\
    SVM & \multicolumn{1}{r|}{2.2333} & SVM & 0.1488 \\ \hline \hline
    \multicolumn{4}{c}{Simulation III for the financial assets dataset} \\ \hline
    $q$ & \multicolumn{1}{l|}{RMSE} & $h$ & \multicolumn{1}{l}{Brier score} \\ \hline
    Multiple regression & \multicolumn{1}{r|}{55827.4190} & Logistic regression & 0.2014 \\
    Random forests & \multicolumn{1}{r|}{58048.7840} & Random forests & 0.2151 \\
    K-nearest neighbor & \multicolumn{1}{r|}{57861.6099} & K-nearest neighbor & 0.2339 \\
    LGBM & \multicolumn{1}{r|}{56665.7031} & LGBM & 0.2047 \\
    SVM & \multicolumn{1}{r|}{65450.8516} & SVM & 0.2080 \\ \hline \hline
    \multicolumn{4}{c}{Simulation III for the jobs dataset} \\ \hline
    $q$ & \multicolumn{1}{l|}{RMSE} & $h$ & \multicolumn{1}{l}{Brier score} \\ \hline
    Multiple regression & \multicolumn{1}{r|}{10963.5250} & Logistic regression & 0.1699 \\
    Random forests & \multicolumn{1}{r|}{12058.0681} & Random forests & 0.1628 \\
    K-nearest neighbor & \multicolumn{1}{r|}{11981.2699} & K-nearest neighbor & 0.1751 \\
    LGBM & \multicolumn{1}{r|}{11503.7386} & LGBM & 0.1719 \\
    SVM & \multicolumn{1}{r|}{15613.6765} & SVM & 0.1912 \\ \hline \hline
    \end{tabular}
\caption{The results of the preliminary simulations.}
\label{tab:preexp}
\end{table}

\section{Settings for the existing methods}
\label{app:methods_settings}
To conduct the exsting method, we used EconML (V0.14.1), FedCI (\url{https://github.com/vothanhvinh/FedCI}), CausalRFF(\url{https://github.com/vothanhvinh/CausalRFF}), FIM and DC-QE.
All hyperparameters of GRF (CausalForest), FedCI and CausalRFF are default values as follows.
\begin{description}
    \item[CausalForest (EconML)] criterion: mse, fit\_intercept: True, honest: True, inference: True, max\_depth: None, max\_features: auto, max\_samples: 0.45, min\_balancedness\_tol: 0.45, min\_impurity\_decrease: 0.0, min\_samples\_leaf: 5, min\_samples\_split: 10, min\_var\_fraction\_leaf: None, min\_var\_leaf\_on\_val: False, min\_weight\_fraction\_leaf: 0.0, n\_estimators: 100, subforest\_size: 4, warm\_start: False
    \item[FedCI] n\_iterations:2000, learning\_rate:1e-3
    \item[CausalRFF] training\_iter: 10000, D: 400, learning\_rate\_w: 1e-2, learning\_rate\_y: 1e-2, learning\_rate\_zy: 1e-1, reg\_beta\_w: 1e-2, reg\_beta\_y: 1e-1, reg\_sig\_y: 1e-1, reg\_beta\_zy: 1e-1, transfer\_flag: FLAGS\_LEARN\_TRANSFER, z\_dim: 80, is\_binary\_outcome: False, use\_mh: True
    \item[FIM] no hyperparameter
    \item[DC-QE] intermediate representation construction method: PCA, $\widetilde{m}_k: m-1$ for all $k$, $\check{m}: m$, propensity score estimation method: inverse probability weighting
\end{description}
For X-Learner, candidate methods for the estimation of the outcome and propensity score have the hyperparameters described in \ref{app:hparams_cand}.
For SR, we considered the regression model to consist of a constant term and a cross term between the covariates and the treatment as
\begin{equation*}
    y_i = \textup{const.} + z_i(\beta^0 + x_i^1 \beta^1 + \cdots + x_i^m \beta^m) + \varepsilon_i.
\end{equation*}

\section{The all results of Simulation III}
\label{app:expIII}

In this section, we include all results of Simulation III.

\begin{figure}[tb]
  \centering
  \includegraphics[width=0.8\linewidth]{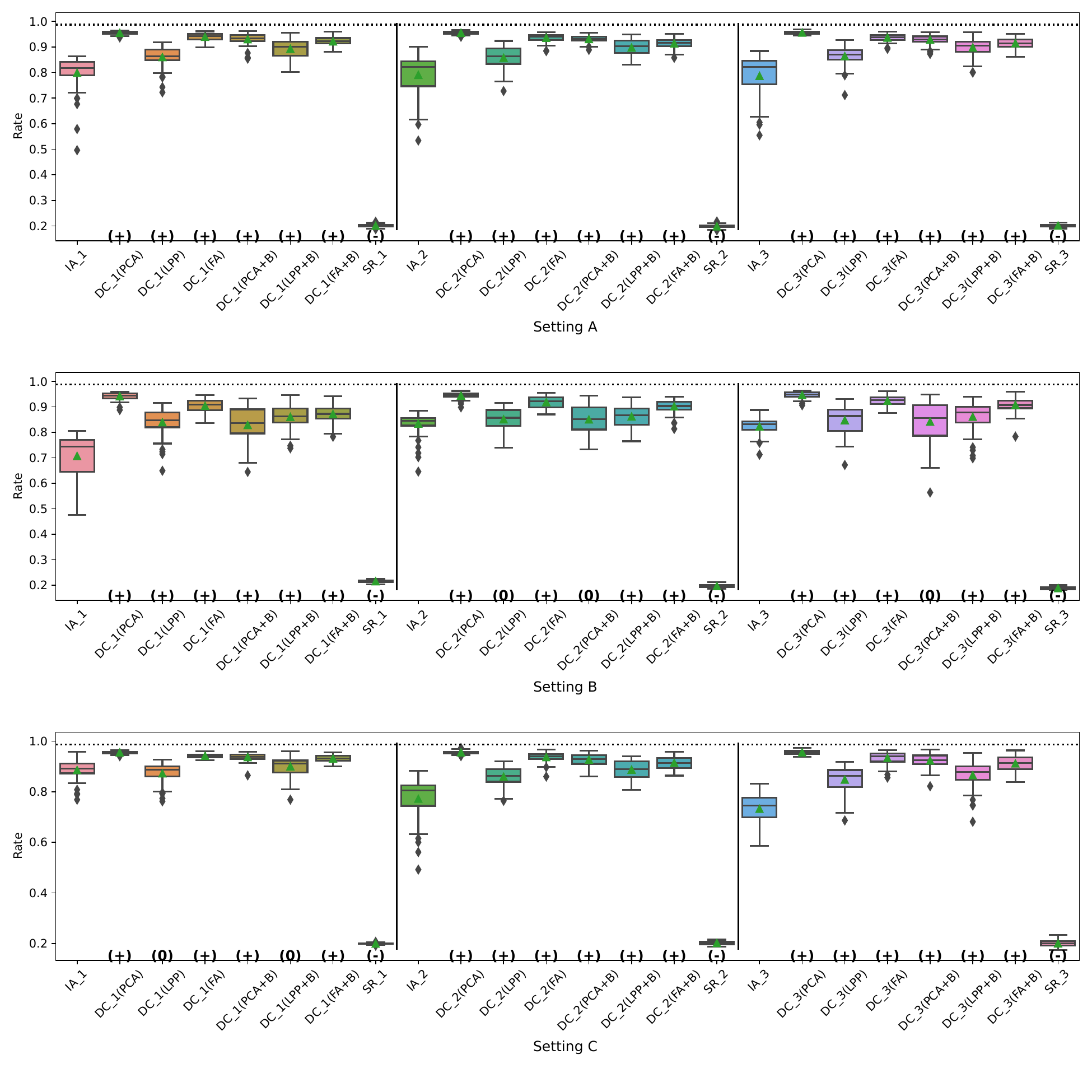}
  \caption{The consistency rate of significance of CATE in Simulation III for the financial assets dataset.}
  \label{fig:exp3_sipp_predsign}
\end{figure}

\begin{figure}[tb]
  \centering
  \includegraphics[width=0.8\linewidth]{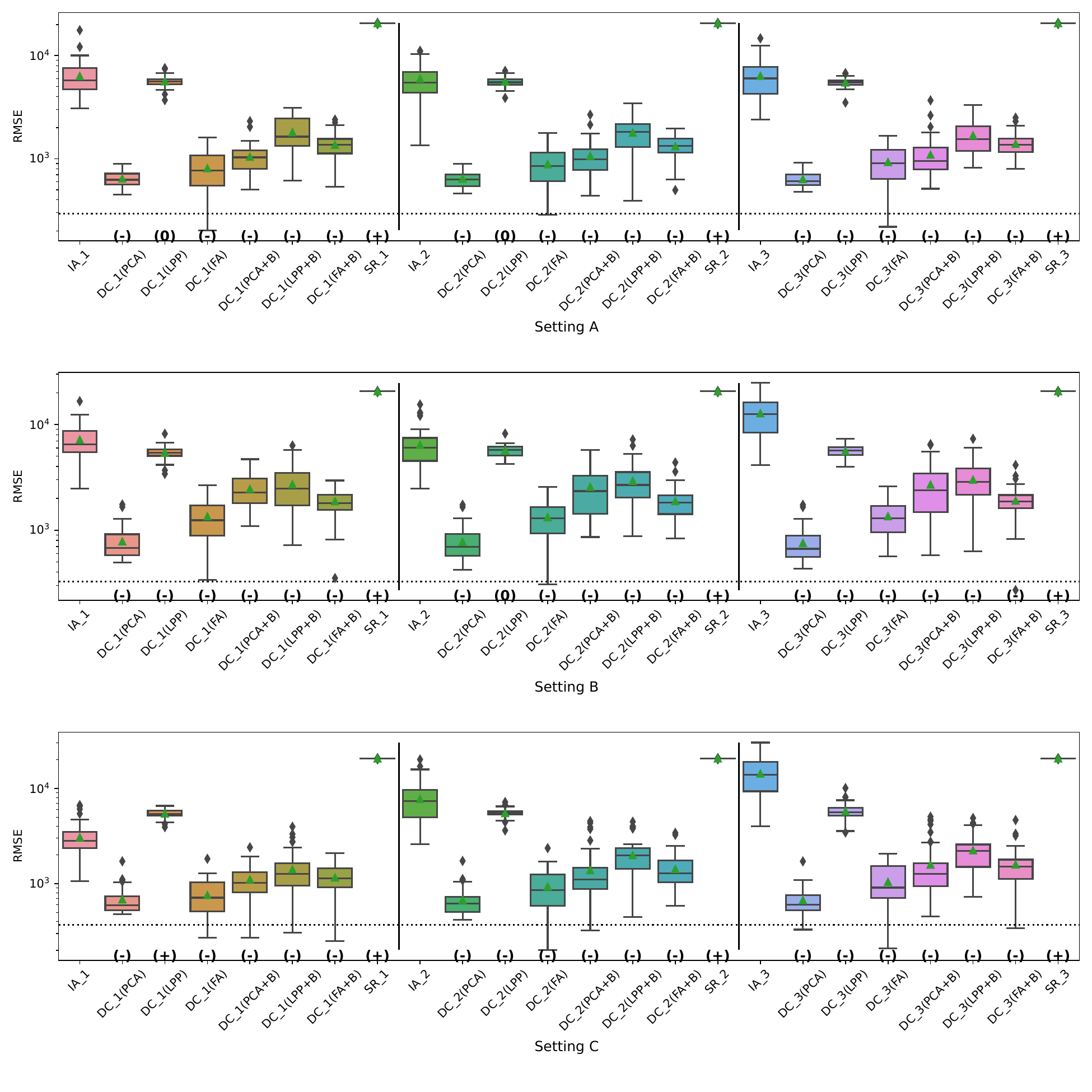}
  \caption{The RMSE of coefficients in Simulation III for the financial assets dataset.}
  \label{fig:exp3_sipp_coefrmse}
\end{figure}

\begin{figure}[tb]
  \centering
  \includegraphics[width=0.8\linewidth]{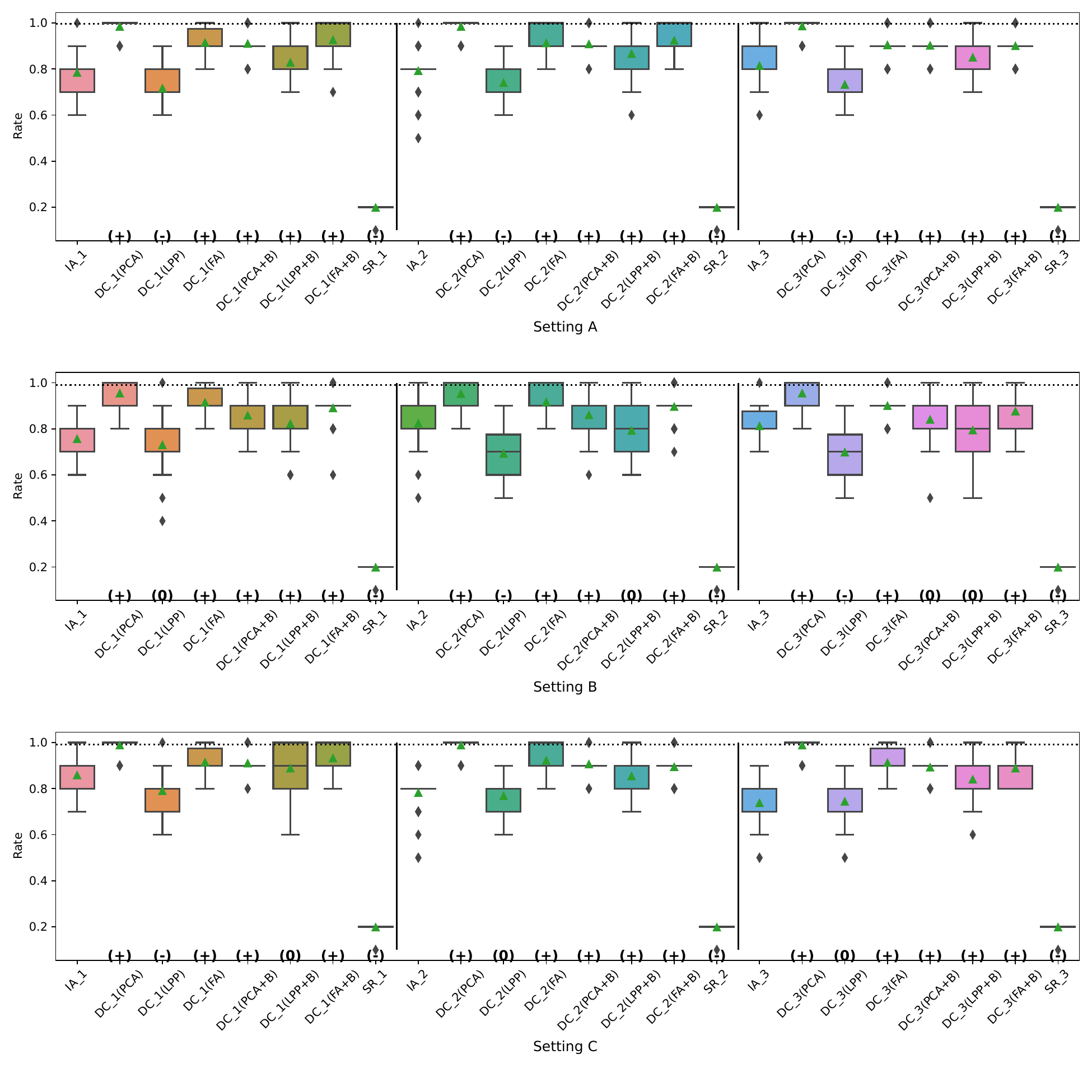}
  \caption{The consistency rate of significance of coefficients in Simulation III for the financial assets dataset.}
  \label{fig:exp3_sipp_coefsign}
\end{figure}

Fig. \ref{fig:exp3_sipp_predrmse}, \ref{fig:exp3_sipp_predsign}, \ref{fig:exp3_sipp_coefrmse} and \ref{fig:exp3_sipp_coefsign} represent the results of Simulation III for the financial assets dataset.
In the RMSE of coefficients and the consistency rate of significance of coefficients shown in Fig. \ref{fig:exp3_sipp_coefrmse} and \ref{fig:exp3_sipp_coefsign}, DC-DML results for party1 using LPP are worse than for IA-DML in some cases.
However, in most other cases shown in Fig. \ref{fig:exp3_sipp_predrmse}, \ref{fig:exp3_sipp_predsign}, \ref{fig:exp3_sipp_coefrmse} and \ref{fig:exp3_sipp_coefsign}, DC-DML results are better than IA-DML.
In addition, the dimensionality reduction method combined with bootstrapping showed robust results.
No SR results are better than IA-DML.

\begin{figure}[tb]
  \centering
  \includegraphics[width=0.8\linewidth]{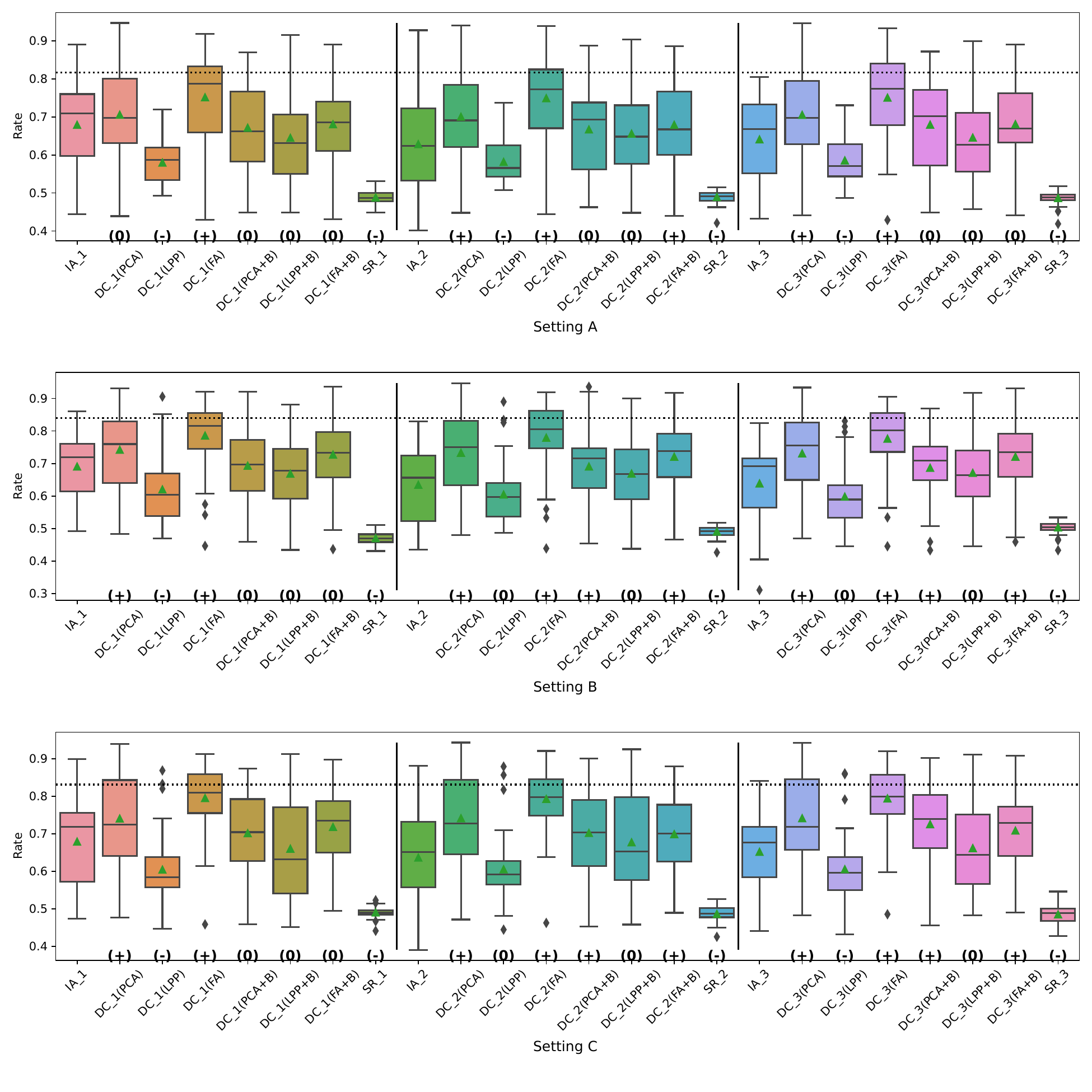}
  \caption{The consistency rate of significance of CATE in Simulation III for the jobs dataset.}
  \label{fig:exp3_jobs_predsign}
\end{figure}

\begin{figure}[tb]
  \centering
  \includegraphics[width=0.8\linewidth]{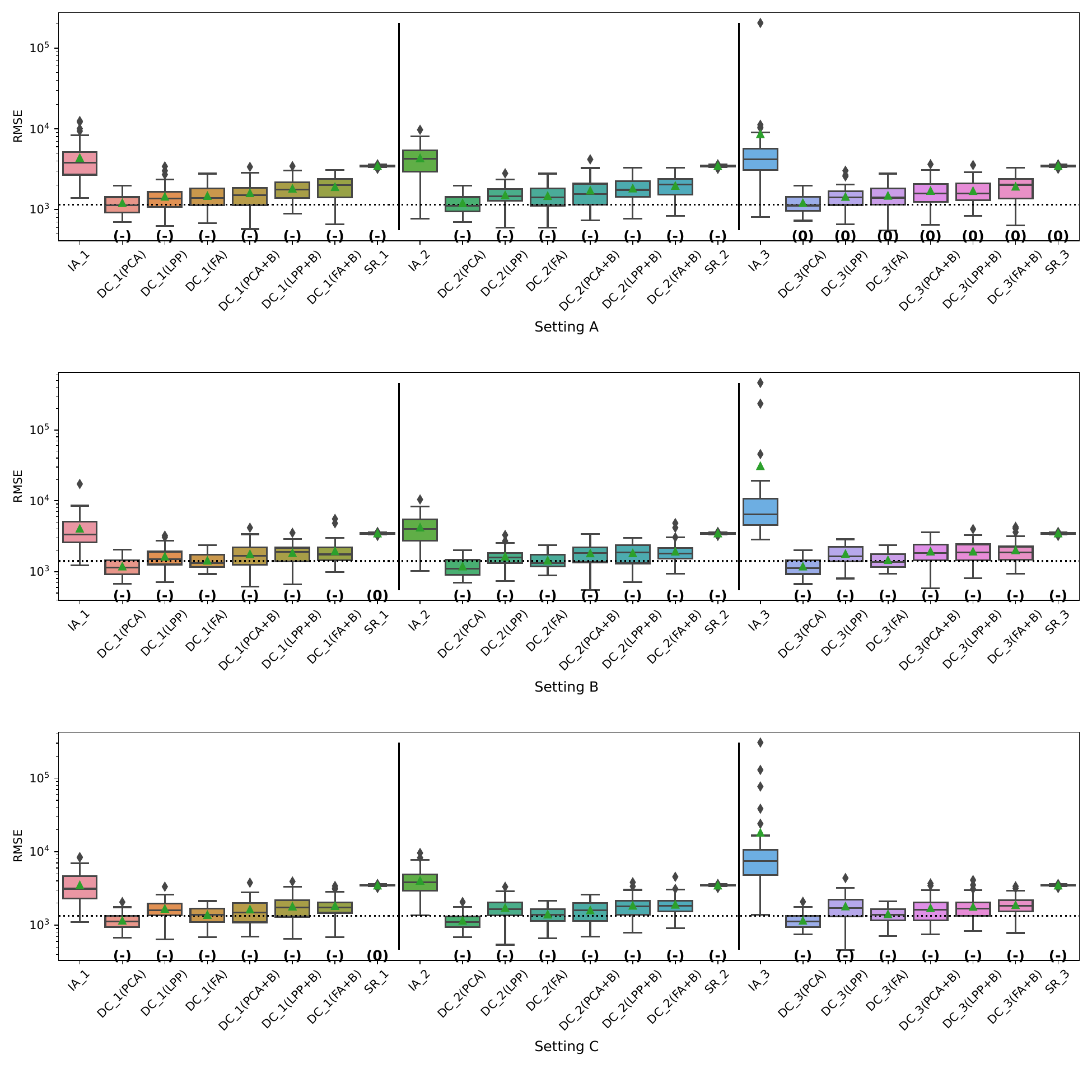}
  \caption{The RMSE of coefficients in Simulation III for the jobs dataset.}
  \label{fig:exp3_jobs_coefrmse}
\end{figure}

\begin{figure}[tb]
  \centering
  \includegraphics[width=0.8\linewidth]{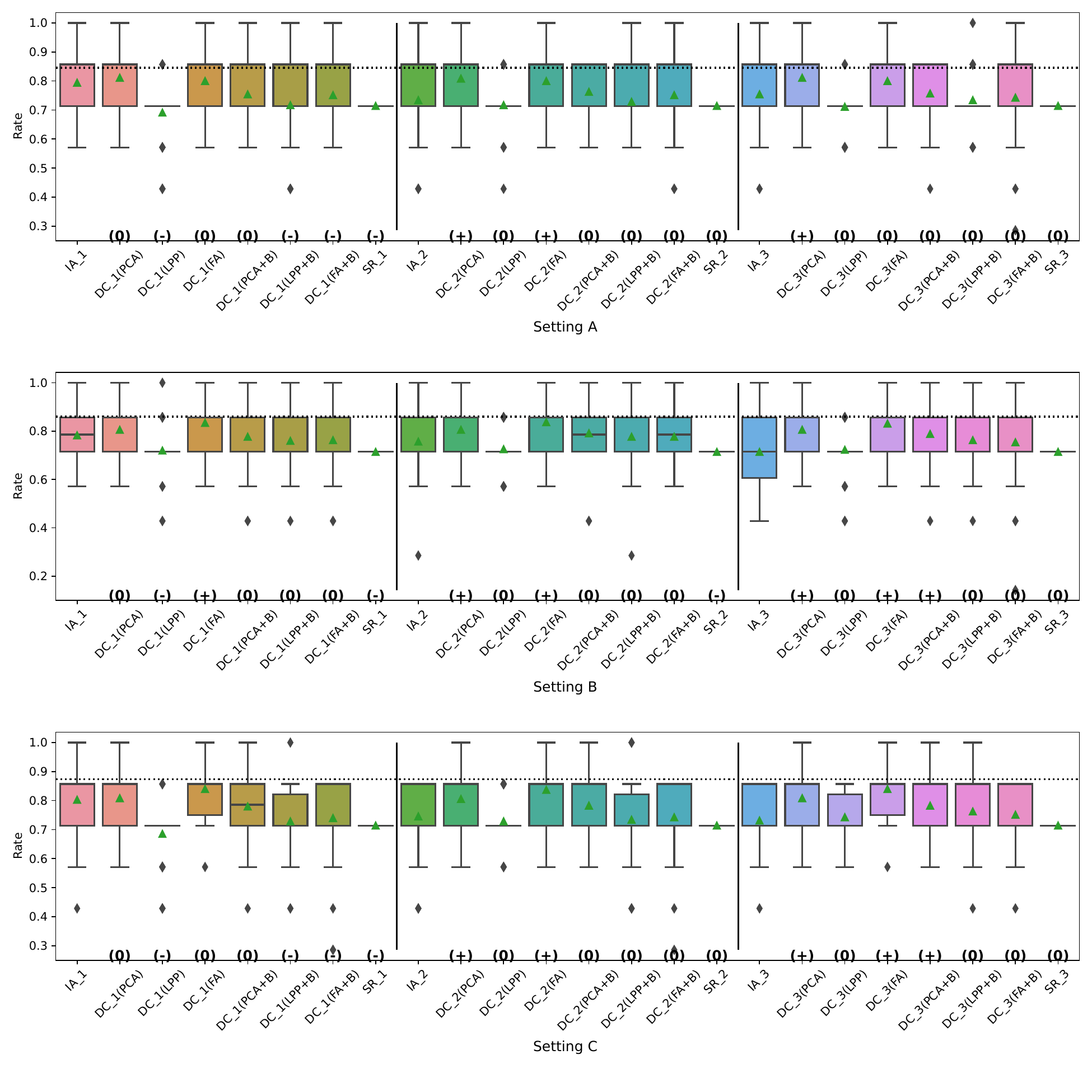}
  \caption{The consistency rate of significance of coefficients in Simulation III for the jobs dataset.}
  \label{fig:exp3_jobs_coefsign}
\end{figure}

Fig. \ref{fig:exp3_jobs_predrmse}, \ref{fig:exp3_jobs_predsign}, \ref{fig:exp3_jobs_coefrmse} and \ref{fig:exp3_jobs_coefsign} represent the results of Simulation III for the jobs dataset.
DC-DML results are worse than IA-DML when using LPP for the consistency rate of significance of CATE, and when using LPP, LPP+B or FA+B for the consistency rate of significance of coefficients.
However, DC-DML results are often better than IA-DML.
DC-DML performs better than IA-DML particularly in the RMSEs of CATE and coefficients in most cases.
Moreover, as with the financial assets dataset, the dimensionality reduction method combined with bootstrapping showed robust results.
SR results are often worse than IA-DML.

In both datasets, there were cases where better performance was not achieved with DC-DML using LPP, especially.
This suggests that the choice of dimensionality reduction method has a significant impact on the performance of DC-DML.

\section{Non-readily identifiable method}
\label{app:non-ri}
The definition of the readily identifiability \citep{imakura2023non} is the following.
\begin{dfn}
    Let $x_i^\textup{p}$ and $x_i^\textup{np}$ be a pair of data that includes and does not include personal information that can directly identify a specific subject for the $i$th person, respectively.
    We let $\mathscr{X}^\textup{p} = \{ x_1^\textup{p}, \cdots, x_n^\textup{p} \}$ and $\mathscr{X}^\textup{np} = \{ x_1^\textup{np}, \cdots, x_n^\textup{np} \}$ be personal information and non-personal information datasets for the same $n$ persons, respectively.

    For non-personal information $x^\textup{np} \in \mathscr{X}^\textup{np}$, if and only if someone else holds a key to correctly collate the corresponding personal information $x^\textup{p} \in \mathscr{X}^\textup{p}$ or can generate the key on their own, then the non-personal dataset $x^\textup{np}$ is defined as ``readily identifiable'' to personal dataset $\mathscr{X}^\textup{p}$.
\end{dfn}
Imakura et al. \citep{imakura2023non} showed that the intermediate representation is readily identifiable to the original dataset in DC analysis.
This is also true for DC-DML, that is, $\widetilde{X}_{k}$ is readily identifiable to $X_{k}$.

To solve the readily identifiable issue in DC analysis, Imakura et al. \citep{imakura2023non} proposed strategies in which the intermediate representations are non-readily identifiable to the original dataset.
To develop DC-DML not to be readily identifiable, the following strategies are needed.
\begin{itemize}
    \item The sample indices in each user are permuted randomly.
    \item Dimensionality reduction functions $f_k$ are erasable and cannot be reconstructed.
\end{itemize}

According to the above strategies, we develop DC-DML as a non-readily identifiable DC-DML (NI-DC-DML) as follows.
In the second stage, the construction of collaborative representations, there are three different procedures between DC-DML and NI-DC-DML.
First, in NI-DC-DML, user $k$ constructs $F_k^\prime = F_k E_k$ instead of $F_k$ as dimensionality reduction function, where $E_k \in \mathbb{R}^{\widetilde{m}_k \times \widetilde{m}_k}$ is a random matrix.
Second, user $k$ constructs a permutation matrix $P_k$.
Thus, the intermediate representations are constructed as
\begin{align*}
    \widetilde{X}_{k} &= P_k (X_{k} - \boldsymbol{1} \boldsymbol{\mu}_k^\top) F_k^\prime \in \mathbb{R}^{n_k \times \widetilde{m}_k}, \\
    \widetilde{X}_{k}^\text{anc} &= (X^\text{anc} - \boldsymbol{1} \boldsymbol{\mu}_k^\top) F_k^\prime \in \mathbb{R}^{r \times \widetilde{m}_k}.
\end{align*}
Note that $P_k$ is not applied for the construction of $\widetilde{X}_{k}^\text{anc}$.
Then, user $k$ erases $P_k$ and $F_k^\prime$.
$P_k$ and $F_k^\prime$ cannot be reconstructed if the random number generator is used with different random seeds for each user.
Third, user $k$ shares the permutated treatments $Z_k^\prime = P_k Z_{k}$ and outcomes $Y_k^\prime = P_k Y_{k}$, instead of $Z_{k}$ and $Y_{k}$, respectively, with the analyst.
Then, user $k$ erases $\widetilde{X}_{k}$, $\widetilde{X}_{k}^\text{anc}$, $Z_k$ and $Y_k$.

In the final stage, estimation of the users, there are two different procedures between DC-DML and NI-DC-DML.
First, the analyst returns $R_k^{\text{Point-anc}}$ and $R_k^{\text{Var-anc}}$ instead of $R_k^{\text{Point}}$ and $R_k^{\text{Var}}$, respectively, where
\begin{align*}
    R_k^{\text{Point-anc}} &= \check{X}_{k}^\text{anc} \check{\boldsymbol{\gamma}}, \\
    R_k^{\text{Var-anc}} &= \check{X}_{k}^\text{anc} \text{Var}(\check{\boldsymbol{\gamma}}) (\check{X}_{k}^\text{anc})^\top, \\
    \check{X}_{k}^\text{anc} &= [\boldsymbol{1} , \widetilde{X}_{k}^\text{anc}] G_k.
\end{align*}
Second, user $k$ calculates $\boldsymbol{\gamma}_k$ and $\text{Var}(\boldsymbol{\gamma}_k)$ as
\begin{align*}
    \boldsymbol{\gamma}_k &=  ((\bar{X}^\text{anc})^\top \bar{X}^\text{anc})^{-1} (\bar{X}^\text{anc})^\top R_k^{\text{Point-anc}} \\
    (&= \bar{F_k} G_k \check{\boldsymbol{\gamma}}), \\
    \text{Var}(\boldsymbol{\gamma}_k) &= ((\bar{X}^\text{anc})^\top \bar{X}^\text{anc})^{-1} (\bar{X}^\text{anc})^\top R_k^{\text{Var-anc}} \bar{X}^\text{anc} ((\bar{X}^\text{anc})^\top \bar{X}^\text{anc})^{-1} \\
    (&= \bar{F_k} G_k \text{Var}(\check{\boldsymbol{\gamma}}) G_k^\top \bar{F_k}^\top), \\
    \bar{X}^\text{anc} &= [\boldsymbol{1} , X^\text{anc}].
\end{align*}
The above calculation is correct because
\begin{align*}
    &((\bar{X}^\text{anc})^\top \bar{X}^\text{anc})^{-1} (\bar{X}^\text{anc})^\top \check{X}_{k}^\text{anc} \\
    &= ((\bar{X}^\text{anc})^\top \bar{X}^\text{anc})^{-1} (\bar{X}^\text{anc})^\top [\boldsymbol{1} , \widetilde{X}_{k}^\text{anc}] G_k \\
    &= ((\bar{X}^\text{anc})^\top \bar{X}^\text{anc})^{-1} (\bar{X}^\text{anc})^\top [\boldsymbol{1} , X^\text{anc}] \bar{F_k} G_k \\
    &= ((\bar{X}^\text{anc})^\top \bar{X}^\text{anc})^{-1} (\bar{X}^\text{anc})^\top \bar{X}^\text{anc} \bar{F_k} G_k \\
    &= \bar{F_k} G_k.
\end{align*}
The other procedures in NI-DC-DML are the same as those in DC-DML.
Algorithm \ref{alg:nidcdml} is the pseudo-code of NI-DC-DML.

\begin{algorithm}[tb]
    \caption{Non-readily identifiable data collaboration double machine learning (NI-DC-DML)}
    \label{alg:nidcdml}
    \begin{algorithmic}[1]
        \Statex \textbf{Input: covariates $X$, treatments $Z$, outcomes $Y$}.
        \Statex \textbf{Output: $\boldsymbol{\beta}_k$, $\text{Var}(\boldsymbol{\gamma}_k)$ and $\text{Var}(\alpha_k)$}.
        \vspace{-.2\baselineskip}
        
        \Statex \hrulefill
        \vspace{-.3\baselineskip}
        \Statex \textit{user-$k$-side}
        \vspace{-.5\baselineskip}
        \Statex \hrulefill
        \vspace{-.1\baselineskip}
        \State Generate anchor dataset $X_{k}^\text{anc}$ and share it with all users.
        \State Set $X^\text{anc}$.
        \State Generate $F_{k}^\prime$, $\boldsymbol{\mu}_k$ and $P_k$.
        \State Compute $\widetilde{X}_{k} = P_k (X_{k} - \boldsymbol{1} \boldsymbol{\mu}_k^\top ) F_{k}^\prime$, $\widetilde{X}_{k}^\text{anc} = (X^\text{anc} - \boldsymbol{1} \boldsymbol{\mu}_k^\top) F_{k}^\prime$, $Z_k^\prime = P_k Z_k$ and $Y_k^\prime = P_k Y_k$.
        \State Erase $F_{k}^\prime$ and $P_k$.
        \State Share $[\boldsymbol{1} , \widetilde{X}_{k}]$, $[\boldsymbol{1} , \widetilde{X}_{k}^\text{anc}]$, $Z_k^\prime$ and $Y_k^\prime$ to the analyst and erase them.
        \vspace{-.3\baselineskip}
        
        \Statex \hrulefill
        \vspace{-.3\baselineskip}
        \Statex \textit{analyst-side}
        \vspace{-.5\baselineskip}
        \Statex \hrulefill
        \vspace{-.1\baselineskip}
        \State Get $[\boldsymbol{1} , \widetilde{X}_{k}]$, $[\boldsymbol{1} , \widetilde{X}_{k}^\text{anc}]$, $Z_k^\prime$ and $Y_k^\prime$ for all $k$.
        \State Set $[\boldsymbol{1} , \widetilde{X}_{k}]$ and $[\boldsymbol{1} , \widetilde{X}_{k}^\text{anc}]$.
        \State Compute $G_k$ from $\widetilde{X}_{k}^\text{anc}$ for all $k$ such that $[\boldsymbol{1} , \widetilde{X}_{k}^\text{anc}] G_k \approx [\boldsymbol{1} , \widetilde{X}_{k'}^\text{anc}] G_{k'} ~ (k \neq k')$.
        \State Compute $\check{X}_k = [\boldsymbol{1} , \widetilde{X}_{k}] G_k$ for all $k$.
        \State Set $\check{X}$, $Z^\prime$ and $Y^\prime$.
        \State Compute function $\check{q}$ and $\check{h}$ using $\check{X}$, $Z^\prime$ and $Y^\prime$.
        \State Compute residuals $\hat{\boldsymbol{\eta}}$ and $\hat{\boldsymbol{\zeta}}$ using $\check{q}$, $\check{h}$, $\check{X}$, $Z^\prime$ and $Y^\prime$.
        \State Obtain $\check{\boldsymbol{\gamma}}$ and $\text{Var}(\check{\boldsymbol{\gamma}})$ as the least square solution of (4.1.2) (in our paper).
        \State Return $R_k^{\text{Point-anc}}$ and $R_k^{\text{Var-anc}}$ to all user $k$.
        \vspace{-.3\baselineskip}

        \Statex \hrulefill
        \vspace{-.3\baselineskip}
        \Statex \textit{user-$k$-side}
        \vspace{-.5\baselineskip}
        \Statex \hrulefill
        \vspace{-.1\baselineskip}
        \State Set $R_k^{\text{Point-anc}}$ and $R_k^{\text{Var-anc}}$.
        \State Compute $\boldsymbol{\beta}_k$, $\text{Var}(\boldsymbol{\gamma}_k)$ and $\text{Var}(\alpha_k)$.
        
        \end{algorithmic}
\end{algorithm}

The estimation result obtained in NI-DC-DML when $E_k$ is the identity matrix is equal to that obtained in DC-DML.
Conversely, if $E_k$ is not the identity matrix, the NI-DC-DML and DC-DML estimation results differ.
However, we think the difference is not large.
As an example, in Table \ref{tab:expapp} we show the estimation results of NI-DC-DML in the situation of Simulation I.
The results are correct except that the coefficient of $x^3$ for party 2 is significant.

Strictly, if $y_i$ or $z_i$ are continuous values, NI-DC-DML cannot satisfy the non-readily identifiability.
The solution to this issue could be the use of other privacy preservation technologies such as $k$-anonymization, differential privacy or cryptography.
That is one of the important future issues.
Further investigation of the relationship between the readily identifiability and DC-DML deviates significantly from the purpose of this paper, thus we end our discussion here.

\begin{table}[tb]
\centering
    \begin{tabular}{l|rlrl}
        \hline \hline
        \multirow{2}{*}{Covariates} & \multicolumn{4}{l}{DC-DML(PCA+B)} \\ \cline{2-5} 
         & Party 1 &  & Party 2 &  \\ \hline
        const. & 0.9133 & ** & 0.9772 & ** \\
        $x^1$ & 1.0451 & ** & 0.9593 & ** \\
        $x^2$ & 0.8847 & ** & 1.0420 & ** \\
        $x^3$ & -0.1566 &  & -0.2426 & * \\
        $x^4$ & -0.0559 &  & 0.0028 &  \\
        $x^5$ & 0.0189 &  & 0.0158 &  \\
        $x^6$ & 0.1160 &  & 0.0359 &  \\
        $x^7$ & 0.0598 &  & -0.1224 &  \\
        $x^8$ & -0.0413 &  & 0.0371 &  \\
        $x^9$ & 0.0492 &  & 0.0802 &  \\
        $x^{10}$ & -0.1166 &  & -0.1608 &  \\ \hline \hline
        \multicolumn{3}{l}{** $p<0.01$, ~ * $p<0.05$} & 
    \end{tabular}
\caption{The estimation results of NI-DC-DML in the situation of Simulation I.}
\label{tab:expapp}
\end{table}

\bibliography{ref}

\end{document}